\documentclass[onecolumn]{emulateapj}
\newcommand{\etal}{et al.}
\begin{document}
\title{The Sixth Data Release of the Sloan Digital Sky Survey}

\author{
Jennifer K. Adelman-McCarthy\altaffilmark{\ref{Fermilab}},
Marcel A. Ag\"ueros\altaffilmark{\ref{Columbia},\ref{NSFFellow}},
Sahar S. Allam\altaffilmark{\ref{Fermilab},\ref{Wyoming}},
Carlos Allende Prieto\altaffilmark{\ref{Texas}},
Kurt S. J. Anderson\altaffilmark{\ref{APO},\ref{NMSU}},
Scott F. Anderson\altaffilmark{\ref{Washington}},
James Annis\altaffilmark{\ref{Fermilab}},
Neta A. Bahcall\altaffilmark{\ref{Princeton}},
C.A.L. Bailer-Jones\altaffilmark{\ref{MPIA}},
Ivan K. Baldry\altaffilmark{\ref{LJMU},\ref{JHU}},
J. C. Barentine\altaffilmark{\ref{Texas}},
Bruce A. Bassett\altaffilmark{\ref{SAAO},\ref{CapeTown}},
Andrew C. Becker\altaffilmark{\ref{Washington}},
Timothy C. Beers\altaffilmark{\ref{MSUJINA}},
Eric F. Bell\altaffilmark{\ref{MPIA}},
Andreas A. Berlind\altaffilmark{\ref{NYU}},
Mariangela Bernardi\altaffilmark{\ref{Penn}}, 
Michael R. Blanton\altaffilmark{\ref{NYU}},
John J. Bochanski\altaffilmark{\ref{Washington}},
William N. Boroski\altaffilmark{\ref{Fermilab}},
Jarle Brinchmann\altaffilmark{\ref{Porto}},
J. Brinkmann\altaffilmark{\ref{APO}},
Robert J. Brunner\altaffilmark{\ref{Illinois}},
Tam\'as Budav\'ari\altaffilmark{\ref{JHU}},
Samuel Carliles\altaffilmark{\ref{JHU}},
Michael A. Carr\altaffilmark{\ref{Princeton}},
Francisco J. Castander\altaffilmark{\ref{Barcelona}},
David Cinabro\altaffilmark{\ref{WayneState}},
R. J. Cool\altaffilmark{\ref{Arizona}},
Kevin R. Covey\altaffilmark{\ref{CfA}},
Istv\'an Csabai\altaffilmark{\ref{Eotvos},\ref{JHU}},
Carlos E. Cunha\altaffilmark{\ref{Chicago},\ref{CfCP}},
James R. A. Davenport\altaffilmark{\ref{Washington}},
Ben Dilday\altaffilmark{\ref{ChicagoPhys},\ref{CfCP}},
Mamoru Doi\altaffilmark{\ref{IoaUT}},
Daniel J. Eisenstein\altaffilmark{\ref{Arizona}},
Michael L. Evans\altaffilmark{\ref{Washington}},
Xiaohui Fan\altaffilmark{\ref{Arizona}},
Douglas P. Finkbeiner\altaffilmark{\ref{CfA}},
Scott D. Friedman\altaffilmark{\ref{STScI}},
Joshua A. Frieman\altaffilmark{\ref{Fermilab},\ref{Chicago},\ref{CfCP}},
Masataka Fukugita\altaffilmark{\ref{ICRRUT}},
Boris T. G\"ansicke\altaffilmark{\ref{Warwick}},
Evalyn Gates\altaffilmark{\ref{Chicago}},
Bruce Gillespie\altaffilmark{\ref{APO}},
Karl Glazebrook\altaffilmark{\ref{Swinburne}},
Jim Gray\altaffilmark{\ref{Microsoft}},
Eva K. Grebel\altaffilmark{\ref{Basel},\ref{Heidelberg}},
James E. Gunn\altaffilmark{\ref{Princeton}},
Vijay K. Gurbani\altaffilmark{\ref{Fermilab},\ref{Lucent2}},
Patrick B. Hall\altaffilmark{\ref{York}},
Paul Harding\altaffilmark{\ref{Case}},
Michael Harvanek\altaffilmark{\ref{Lowell}},
Suzanne L. Hawley\altaffilmark{\ref{Washington}},
Jeffrey Hayes\altaffilmark{\ref{Catholic}},
Timothy M. Heckman\altaffilmark{\ref{JHU}},
John S. Hendry\altaffilmark{\ref{Fermilab}},
Robert B. Hindsley\altaffilmark{\ref{NRL}},
Christopher M. Hirata\altaffilmark{\ref{IAS}},
Craig J. Hogan\altaffilmark{\ref{Washington}},
David W. Hogg\altaffilmark{\ref{NYU}},
Joseph B. Hyde\altaffilmark{\ref{Penn}},
Shin-ichi Ichikawa\altaffilmark{\ref{NAOJ}},
\v{Z}eljko Ivezi\'{c}\altaffilmark{\ref{Washington}},
Sebastian Jester\altaffilmark{\ref{MPIA}},
Jennifer A. Johnson\altaffilmark{\ref{OSU}},
Anders M. Jorgensen\altaffilmark{\ref{NMIMT}},
Mario Juri\'{c}\altaffilmark{\ref{IAS}},
Stephen M. Kent\altaffilmark{\ref{Fermilab}},
R. Kessler\altaffilmark{\ref{EFI}}, 
S. J. Kleinman\altaffilmark{\ref{Gemini}},
G. R. Knapp\altaffilmark{\ref{Princeton}},
Richard G. Kron\altaffilmark{\ref{Chicago},\ref{Fermilab}},
Jurek Krzesinski\altaffilmark{\ref{APO},\ref{MSO}},
Nikolay Kuropatkin\altaffilmark{\ref{Fermilab}},
Donald Q. Lamb\altaffilmark{\ref{Chicago},\ref{EFI}},
Hubert Lampeitl\altaffilmark{\ref{STScI}},
Svetlana Lebedeva\altaffilmark{\ref{Fermilab}},
Young Sun Lee\altaffilmark{\ref{MSUJINA}},
R. French Leger\altaffilmark{\ref{Fermilab}},
S\'ebastien L\'epine\altaffilmark{\ref{AMNH}},
Marcos Lima\altaffilmark{\ref{ChicagoPhys},\ref{CfCP}},
Huan Lin\altaffilmark{\ref{Fermilab}},
Daniel C. Long\altaffilmark{\ref{APO}},
Craig P. Loomis\altaffilmark{\ref{Princeton}},
Jon Loveday\altaffilmark{\ref{Sussex}},
Robert H. Lupton\altaffilmark{\ref{Princeton}},
Olena Malanushenko\altaffilmark{\ref{APO}},
Viktor Malanushenko\altaffilmark{\ref{APO}},
Rachel Mandelbaum\altaffilmark{\ref{IAS},\ref{Hubble}},
Bruce Margon\altaffilmark{\ref{SantaCruz}},
John P. Marriner\altaffilmark{\ref{Fermilab}},
David Mart\'{\i}nez-Delgado\altaffilmark{\ref{IAC}},
Takahiko Matsubara\altaffilmark{\ref{Nagoya}},
Peregrine M. McGehee\altaffilmark{\ref{IPAC}},
Timothy A. McKay\altaffilmark{\ref{Michigan}},
Avery Meiksin\altaffilmark{\ref{Edinburgh}},
Heather L. Morrison\altaffilmark{\ref{Case}},
Jeffrey A. Munn\altaffilmark{\ref{NOFS}},
Reiko Nakajima\altaffilmark{\ref{Penn}},
Eric H. Neilsen, Jr.\altaffilmark{\ref{Fermilab}},
Heidi Jo Newberg\altaffilmark{\ref{RPI}},
Robert C. Nichol\altaffilmark{\ref{Portsmouth}},
Tom Nicinski\altaffilmark{\ref{Fermilab},\ref{CMCElectronics}},
Maria Nieto-Santisteban\altaffilmark{\ref{JHU}},
Atsuko Nitta\altaffilmark{\ref{Gemini}},
Sadanori Okamura\altaffilmark{\ref{DoAUT}},
Russell Owen\altaffilmark{\ref{Washington}},
Hiroaki Oyaizu\altaffilmark{\ref{Chicago},\ref{CfCP}},
Nikhil Padmanabhan\altaffilmark{\ref{LBL},\ref{Hubble}},
Kaike Pan\altaffilmark{\ref{APO}},
Changbom Park\altaffilmark{\ref{KIAS}},
John Peoples Jr.\altaffilmark{\ref{Fermilab}},
Jeffrey R. Pier\altaffilmark{\ref{NOFS}},
Adrian C. Pope\altaffilmark{\ref{Hawaii}},
Norbert Purger\altaffilmark{\ref{Eotvos}},
M. Jordan Raddick\altaffilmark{\ref{JHU}},
Paola Re Fiorentin\altaffilmark{\ref{MPIA}},
Gordon T. Richards\altaffilmark{\ref{Drexel}},
Michael W. Richmond\altaffilmark{\ref{RIT}},
Adam G. Riess\altaffilmark{\ref{JHU}},
Hans-Walter Rix\altaffilmark{\ref{MPIA}},
Constance M. Rockosi\altaffilmark{\ref{Lick}},
Masao Sako\altaffilmark{\ref{Penn},\ref{Stanford}},
David J. Schlegel\altaffilmark{\ref{LBL}},
Donald P. Schneider\altaffilmark{\ref{PSU}},
Matthias R. Schreiber\altaffilmark{\ref{Valparaiso}},
Axel D. Schwope\altaffilmark{\ref{Potsdam}},
Uro\v{s} Seljak\altaffilmark{\ref{Princetonphys},\ref{Zurich}},
Branimir Sesar\altaffilmark{\ref{Washington}},
Erin Sheldon\altaffilmark{\ref{Chicago},\ref{CfCP}},
Kazu Shimasaku\altaffilmark{\ref{DoAUT}},
Thirupathi Sivarani\altaffilmark{\ref{MSUJINA}},
J. Allyn Smith\altaffilmark{\ref{APeay}},
Stephanie A. Snedden\altaffilmark{\ref{APO}},
Matthias Steinmetz\altaffilmark{\ref{Potsdam}},
Michael A. Strauss\altaffilmark{\ref{Princeton}},
Mark SubbaRao\altaffilmark{\ref{Chicago},\ref{Adler}},
Yasushi Suto\altaffilmark{\ref{TokyoPhys}},
Alexander S. Szalay\altaffilmark{\ref{JHU}},
Istv\'an Szapudi\altaffilmark{\ref{Hawaii}},
Paula Szkody\altaffilmark{\ref{Washington}},
Max Tegmark\altaffilmark{\ref{MIT}},
Aniruddha R. Thakar\altaffilmark{\ref{JHU}},
Christy A. Tremonti\altaffilmark{\ref{Arizona}},
Douglas L. Tucker\altaffilmark{\ref{Fermilab}},
Alan Uomoto\altaffilmark{\ref{CarnegieObs}},
Daniel E. Vanden Berk\altaffilmark{\ref{PSU}},
Jan Vandenberg\altaffilmark{\ref{JHU}},
S. Vidrih\altaffilmark{\ref{Cambridge}},
Michael S. Vogeley\altaffilmark{\ref{Drexel}},
Wolfgang Voges\altaffilmark{\ref{MPIEP}},
Nicole P. Vogt\altaffilmark{\ref{NMSU}},
Yogesh Wadadekar\altaffilmark{\ref{Princeton}},
David H. Weinberg\altaffilmark{\ref{OSU}},
Andrew A. West\altaffilmark{\ref{Berkeley}},
Simon D.M. White\altaffilmark{\ref{MPA}},
Brian C. Wilhite\altaffilmark{\ref{Illinois},\ref{NCSA}},
Brian Yanny\altaffilmark{\ref{Fermilab}},
D. R. Yocum\altaffilmark{\ref{Fermilab}},
Donald G. York\altaffilmark{\ref{Chicago},\ref{EFI}},
Idit Zehavi\altaffilmark{\ref{Case}},
Daniel B. Zucker\altaffilmark{\ref{Cambridge}}
}

\altaffiltext{1}{
Fermi National Accelerator Laboratory, P.O. Box 500, Batavia, IL 60510.
\label{Fermilab}}

\altaffiltext{2}{
 	Columbia Astrophysics Laboratory, 550 West 120th Street, New
	York, NY 10027.\label{Columbia}}

\altaffiltext{3}{
         NSF Astronomy and Astrophysics
	 Postdoctoral Fellow.\label{NSFFellow}}

\altaffiltext{4}{
Department of Physics and Astronomy, University of Wyoming, Laramie, WY 82071.
\label{Wyoming}}

\altaffiltext{5}{
McDonald Observatory and Department of Astronomy, The University of
Texas, 1 University Station, C1400,  Austin, TX
78712-0259.\label{Texas}}

\altaffiltext{6}{
Apache Point Observatory, P.O. Box 59, Sunspot, NM 88349.
\label{APO}}

\altaffiltext{7}{
Department of Astronomy, MSC 4500, New Mexico State University,
P.O. Box 30001, Las Cruces, NM 88003.
\label{NMSU}}

\altaffiltext{8}{
Department of Astronomy, University of Washington, Box 351580, Seattle, WA
98195.
\label{Washington}}

\altaffiltext{9}{
Department of Astrophysical Sciences, Princeton University, Princeton, NJ
08544.
\label{Princeton}}

\altaffiltext{10}{
Max-Planck-Institut f\"ur Astronomie, K\"onigstuhl 17, D-69117 Heidelberg,
Germany.
\label{MPIA}}

\altaffiltext{11}{
Astrophysics Research Institute,
Liverpool John Moores University,
Twelve Quays House, Egerton Wharf,
Birkenhead CH41 1LD, UK.
\label{LJMU}}

\altaffiltext{12}{
Center for Astrophysical Sciences, Department of Physics and Astronomy, Johns
Hopkins University, 3400 North Charles Street, Baltimore, MD 21218. 
\label{JHU}}

\altaffiltext{13}{
South African Astronomical Observatory, Observatory,
  Cape Town, South Africa.\label{SAAO}}

\altaffiltext{14}{
University of Cape Town, Rondebosch, Cape Town, South
  Africa.
\label{CapeTown}}

\altaffiltext{15}{
Dept. of Physics \& Astrophysics, CSCE: Center for the Study of Cosmic 
Evolution, and JINA: Joint Institute for Nuclear Astrophysics, Michigan 
State University, E. Lansing, MI  48824, USA.
\label{MSUJINA}}

\altaffiltext{16}{
Center for Cosmology and Particle Physics,
Department of Physics,
New York University,
4 Washington Place,
New York, NY 10003.
\label{NYU}}

\altaffiltext{17}{
Department of Physics and Astronomy, University of Pennsylvania,
209 South 33rd Street, Philadelphia, PA 19104. 
\label{Penn}}

\altaffiltext{18}{
Centro de Astrof{\'\i}sica da Universidade do Porto, Rua 
das Estrelas - 4150-762 Porto, Portugal.
\label{Porto}}

\altaffiltext{19}{
Department of Astronomy,
University of Illinois,
1002 West Green Street, Urbana, IL 61801.
\label{Illinois}}

\altaffiltext{20}{
Institut de Ci\`encies de l'Espai (IEEC/CSIC),
  Campus UAB, E-08193  
Bellaterra, Barcelona, Spain.
\label{Barcelona}}

\altaffiltext{21}{
Department of Physics and Astronomy, Wayne State University,
             Detroit, MI 48202.\label{WayneState}}

\altaffiltext{22}{
Steward Observatory, 933 North Cherry Avenue, Tucson, AZ 85721.
\label{Arizona}}

\altaffiltext{23}{
Harvard-Smithsonian Center for Astrophysics, 60
  Garden Street, Cambridge MA 02138.\label{CfA}}

\altaffiltext{24}{
Department of Physics of Complex Systems, 
E\"{o}tv\"{o}s Lor\'and University, Pf.\ 32,
H-1518 Budapest, Hungary.
\label{Eotvos}}

\altaffiltext{25}{
Department of Astronomy and Astrophysics, University of Chicago, 5640 South
Ellis Avenue, Chicago, IL 60637.
\label{Chicago}}

\altaffiltext{26}{
Kavli Institute for Cosmological Physics, The University of Chicago,
5640 South Ellis Avenue, Chicago, IL 60637.
\label{CfCP}}

\altaffiltext{27}{
Department of Physics, University of Chicago, 5640 South
Ellis Avenue, Chicago, IL 60637.
\label{ChicagoPhys}}

\altaffiltext{28}{
Institute of Astronomy,
Graduate School of Science, The University of Tokyo,
2-21-1 Osawa, Mitaka, 181-0015, Japan.
\label{IoaUT}}

\altaffiltext{29}{
Space Telescope Science Institute, 3700 San Martin Drive, Baltimore, MD
21218.
\label{STScI}}

\altaffiltext{30}{
Institute for Cosmic Ray Research, The University of Tokyo, 5-1-5 Kashiwa,
 Kashiwa City, Chiba 277-8582, Japan.
\label{ICRRUT}}

\altaffiltext{31}{
Department of Physics, University of Warwick,
Coventry CV4 7AL, United Kingdom.\label{Warwick}}

\altaffiltext{32}{
Centre for Astrophysics \& Supercomputing, Swinburne
   University of Technology, P.O. Box 218, Hawthorn, VIC 3122,
   Australia.\label{Swinburne}}

\altaffiltext{33}{
Microsoft Research, 455 Market Street, Suite 1690, San Francisco, CA 94105.
\label{Microsoft}}

\altaffiltext{34}{
Astronomical Institute, Department of Physics and Astronomy,
University of Basel, Venusstrasse 7, CH-4102 Binningen,
Switzerland.
\label{Basel}}

\altaffiltext{35}{
Astronomisches Rechen-Institut, Zentrum f\"ur Astronomie,
University of Heidelberg, M\"onchhofstrasse 12-14,
D-69120 Heidelberg, Germany.\label{Heidelberg}}

\altaffiltext{36}{ 
Bell Laboratories, Alcatel-Lucent, 2701 Lucent Lane, Rm.~9F-546,
Lisle, Illinois 60532. \label{Lucent2}}

\altaffiltext{37}{
Dept. of Physics \& Astronomy,
York University,
4700 Keele St.,
Toronto, ON, M3J 1P3,
Canada
\label{York}}

\altaffiltext{38}{
Department of Astronomy, Case Western Reserve University,
Cleveland, OH 44106.
\label{Case}}

\altaffiltext{39}{
Lowell Observatory, 
1400 W Mars Hill Rd, 
Flagstaff AZ 86001.\label{Lowell}}

\altaffiltext{40}{
Institute for Astronomy and Computational Sciences, 
     Physics Department, 
     Catholic University of America, 
     Washington DC 20064
\label{Catholic}}

\altaffiltext{41}{
Code 7215, Remote Sensing Division,
Naval Research Laboratory, 
4555 Overlook Avenue SW,
Washington, DC 20392.
\label{NRL}}

\altaffiltext{42}{
Institute for Advanced Study,
Einstein Drive,
Princeton, NJ 08540.
\label{IAS}}

\altaffiltext{43}{
National Astronomical Observatory, 
2-21-1 Osawa, Mitaka, Tokyo 181-8588, Japan.
\label{NAOJ}}

\altaffiltext{44}{
Department of Astronomy, 
Ohio State University, 140 West 18th Avenue, Columbus, OH 43210.
\label{OSU}}

\altaffiltext{45}{
Electrical Engineering Department,
New Mexico Institute of Mining and Technology,
801 Leroy Place,
Socorro, NM 87801.\label{NMIMT}}

\altaffiltext{46}{
Enrico Fermi Institute, University of Chicago, 5640 South Ellis Avenue,
Chicago, IL 60637.
\label{EFI}}

\altaffiltext{47}{
Gemini Observatory, 670 N. A'ohoku Place, Hilo, HI 96720, USA
\label{Gemini}}

\altaffiltext{48}{
Obserwatorium Astronomiczne na Suhorze, Akademia Pedogogiczna w
Krakowie, ulica Podchor\c{a}\.{z}ych 2,
PL-30-084 Krac\'ow, Poland.
\label{MSO}}

\altaffiltext{49}{
  Department of Astrophysics,
American Museum of Natural History,
Central Park West at 79th Street,
New York, NY 10024\label{AMNH}}

\altaffiltext{50}{
Astronomy Centre, University of Sussex, Falmer, Brighton BN1 9QH, UK. 
\label{Sussex}}

\altaffiltext{51}{
Hubble Fellow.\label{Hubble}}

\altaffiltext{52}{
Department of Astronomy \& Astrophysics, University of California, Santa
Cruz, CA 95064.
\label{SantaCruz}}

\altaffiltext{53}{
Instituto de Astrofisica de Canarias, La Laguna, Spain.
\label{IAC}}

\altaffiltext{54}{
Department of Physics and Astrophysics,
 Nagoya University,
 Chikusa, Nagoya 464-8602,
 Japan.
\label{Nagoya}}

\altaffiltext{55}{
IPAC, MS 220-6, California Institute of Technology,
Pasadena, CA 91125.\label{IPAC}}

\altaffiltext{56}{
Department of Physics, University of Michigan, 500 East University Avenue, Ann
Arbor, MI 48109.
\label{Michigan}}

\altaffiltext{57}{
SUPA, Institute for Astronomy,
Royal Observatory,
University of Edinburgh,
Blackford Hill,
Edinburgh EH9 3HJ,
UK.
\label{Edinburgh}}

\altaffiltext{58}{
US Naval Observatory, 
Flagstaff Station, 10391 W. Naval Observatory Road, Flagstaff, AZ
86001-8521.
\label{NOFS}}

\altaffiltext{59}{
Department of Physics, Applied Physics, and Astronomy, Rensselaer
Polytechnic Institute, 110 Eighth Street, Troy, NY 12180. 
\label{RPI}}

\altaffiltext{60}{
Institute of Cosmology and Gravitation (ICG),
Mercantile House, Hampshire Terrace,
Univ. of Portsmouth, Portsmouth, PO1 2EG, UK.
\label{Portsmouth}}

\altaffiltext{61}{
    CMC Electronics Aurora,
 84 N. Dugan Rd.
    Sugar Grove, IL 60554.
\label{CMCElectronics}}

\altaffiltext{62}{
Department of 
Astronomy and Research Center for the Early Universe, 
Graduate School of Science, The University of Tokyo,
 7-3-1 Hongo, Bunkyo, Tokyo 113-0033, Japan.
\label{DoAUT}}

\altaffiltext{63}{
Lawrence Berkeley National Laboratory, One Cyclotron Road,
Berkeley, CA 94720.
\label{LBL}}

\altaffiltext{64}{
Korea Institute for Advanced Study,
207-43 Cheong-Nyang-Ni, 2 dong,
Seoul 130-722, Korea
\label{KIAS}}

\altaffiltext{65}{
Institute for Astronomy, 2680 Woodlawn Road, Honolulu, HI 96822.
\label{Hawaii}}

\altaffiltext{66}{
Department of Physics, 
Drexel University, 3141 Chestnut Street, Philadelphia, PA 19104.
\label{Drexel}}

\altaffiltext{67}{
Department of Physics, Rochester Institute of Technology, 84 Lomb Memorial
Drive, Rochester, NY 14623-5603.
\label{RIT}}

\altaffiltext{68}{
UCO/Lick Observatory, University of California, Santa Cruz, CA 95064.
\label{Lick}}

\altaffiltext{69}{
   Kavli Institute for Particle Astrophysics \& Cosmology,
   Stanford University, P.O. Box 20450, MS29,
   Stanford, CA 94309.\label{Stanford}}

\altaffiltext{70}{
Department of Astronomy and Astrophysics, 525 Davey Laboratory, 
Pennsylvania State
University, University Park, PA 16802.
\label{PSU}}

\altaffiltext{71}{
Universidad de Valparaiso, Departamento de Fisica y
  Astronomia,
 Valparaiso, Chile.\label{Valparaiso}}

\altaffiltext{72}{
Astrophysical Institute Potsdam, An der Sternwarte 16, 
14482 Potsdam, Germany.
\label{Potsdam}}

\altaffiltext{73}{
Joseph Henry Laboratories, Princeton University, Princeton, NJ
08544.
\label{Princetonphys}}

\altaffiltext{74}{
Institute for Theoretical Physics, University of Zurich, Zurich 8057  
Switzerland.\label{Zurich}}

\altaffiltext{75}{
Department of Physics and Astronomy, Austin Peay State University,
P.O. Box 4608, Clarksville, TN 37040.
\label{APeay}}

\altaffiltext{76}{
Adler Planetarium and Astronomy Museum,
1300 Lake Shore Drive,
Chicago, IL 60605.
\label{Adler}}

\altaffiltext{77}{
Department of Physics and Research Center for the Early Universe,
Graduate School of Science, The University of Tokyo, 7-3-1 Hongo, Bunkyo, 
Tokyo 113-0033, Japan.
\label{TokyoPhys}}

\altaffiltext{78}{
Dept. of Physics, Massachusetts Institute of Technology, Cambridge,  
MA 02139.
\label{MIT}}

\altaffiltext{79}{
Observatories of the Carnegie Institution of Washington, 
813 Santa Barbara Street, 
Pasadena, CA  91101.
\label{CarnegieObs}}

\altaffiltext{80}{
Institute of Astronomy, University of Cambridge, Madingley Road,
Cambridge CB3 0HA, UK.
\label{Cambridge}}

\altaffiltext{81}{
Max-Planck-Institut f\"ur extraterrestrische Physik, 
Giessenbachstrasse 1, D-85741 Garching, Germany.
\label{MPIEP}}

\altaffiltext{82}{
Astronomy Department, 601 Campbell Hall, University of
California, Berkeley, CA 94720-3411.
\label{Berkeley}}

\altaffiltext{83}{
Max Planck Institut f\"ur Astrophysik, Postfach 1, 
D-85748 Garching, Germany.
\label{MPA}}

\altaffiltext{84}{
National Center for Supercomputing Applications,
1205 West Clark Street,
Urbana, IL 61801.\label{NCSA}}

\shorttitle{SDSS DR6}
\shortauthors{Adelman-McCarthy \etal}

\begin{abstract}
This paper describes the Sixth Data Release of the Sloan Digital Sky
Survey.  With this data release, the imaging of the Northern Galactic
Cap is now complete.  The survey contains images and parameters of
roughly 287 million objects over 9583 deg$^2$, including scans over a
large range of Galactic latitudes and longitudes.  The survey also
includes 1.27 million spectra of stars, galaxies, quasars and blank
sky (for sky subtraction) selected over 7425 deg$^2$.  This release
includes much more extensive stellar spectroscopy than previously, and
also includes detailed estimates of stellar temperatures, gravities,
and metallicities.  The results of improved photometric calibration
are now available, with uncertainties of roughly 1\% in $g, r, i$, and
$z$, and 2\% in $u$, substantially better than the uncertainties in
previous data releases.  The spectra in this data release have
improved wavelength and flux calibration, especially in the extreme
blue and extreme red, leading to the qualitatively better
determination of stellar types and radial velocities.  The
spectrophotometric fluxes are now tied to point spread function
magnitudes of stars rather than fiber magnitudes.  This gives more
robust results in the presence of seeing variations, but also implies
a change in the spectrophotometric scale, which is now brighter by
roughly 0.35 mags.  Systematic errors in the velocity dispersions of
galaxies have been fixed, and the results of two independent codes for
determining spectral classifications and redshifts are made available.
Additional spectral outputs are made available, including calibrated
spectra from individual 15-minute exposures and the sky spectrum
subtracted from each exposure.  We also quantify a recently recognized
under-estimation of the brightnesses of galaxies of large angular
extent due to poor sky subtraction; the bias can exceed 0.2 mag for
galaxies brighter than $r=14$ mag.
\end{abstract}
\section{Introduction}
\label{sec:introduction}
The Sloan Digital Sky Survey (SDSS; York \etal\ 2000) is a comprehensive imaging and
spectroscopic survey of the optical sky using a dedicated 2.5-meter
telescope (Gunn \etal\ 2006) at Apache Point Observatory in southern
New Mexico.  The telescope has a $3^\circ$ diameter field of view, and
the imaging uses a drift-scanning camera (Gunn \etal\ 1998) with 30 $2048 \times
2048$ CCDs at the focal plane which image the sky in five broad
filters covering the range from 3000\AA\ to 10,000\AA\ 
(Fukugita \etal\ 1996; Stoughton \etal\ 2002).  The imaging is carried out on moonless
and cloudless nights of good seeing (Hogg \etal\ 2001), and the resulting
images are calibrated photometrically (Tucker \etal\ 2006; Padmanabhan
\etal\ 2007) to a series of photometric standards around the sky
(Smith \etal\ 2002).  After astrometric calibration (Pier \etal\ 2003)
the properties of detected objects in the five filters are measured in
detail (Lupton \etal\ 2001; Stoughton \etal\ 2002).  Subsets of these
objects are selected for spectroscopy, including galaxies (Strauss
\etal\ 2002; Eisenstein \etal\ 2001), quasars (Richards \etal\ 2002),
and stars.  The spectroscopic targets are assigned to a series of plates
containing 640 objects each (Blanton \etal\ 2003), and spectra are
measured using a pair of double spectrographs, each covering the
wavelength range 3800--9200\AA\ with a resolution $\lambda/\Delta
\lambda$ which varies from 1850 to 2200. 
These spectra are wavelength- and flux-calibrated,
and classifications and redshifts, as well as spectral types
for stars, are determined by a series of software pipelines
(Subbarao \etal\ 2002). The data are then made available both through
an object-oriented database (the Catalog Archive Server, hereafter
``CAS''), and as flat data files (the Data Archive Server, hereafter
``DAS'').  

  The SDSS telescope saw first light in May 1998, and entered routine
  operations in April 2000.  We have issued a series of yearly public
  data releases, which have been described in accompanying papers
  (Stoughton \etal\ 2000, hereafter the Early Data Release, or EDR
  paper; Abazajian \etal\ 2003, 2004, 2005; hereafter the DR1, DR2,
  and DR3 papers respectively, and Adelman-McCarthy \etal\ 2006, 2007;
  hereafter the DR4 and DR5 papers, respectively).  The current paper describes the
  Sixth Data Release (DR6), which includes data taken through June
  2006.  Access to the data themselves may be found on the DR6
  website\footnote{\url{\tt http://www.sdss.org/dr6}}.  This website
  includes links to both the CAS and DAS websites, which contain
  extensive documentation on how to access the data. 

  When the SDSS started routine operations, the budget funded
  operations for five years, i.e., through summer 2005.  Additional
  funding from the National Science Foundation, the Alfred P. Sloan
  Foundation, and the member institutions secured another three years
  of operations, and the present data release includes data from the
  first year of this extended period, termed SDSS-II.  SDSS-II
  has three components: {\em Legacy}, which aims to complete the
  imaging and spectroscopy of a contiguous $\sim 7700$ deg$^2$ region
  in the Northern Galactic Cap, {\em SEGUE} (Sloan Extension for
  Galactic Understanding and Exploration), which is carrying out an
  additional 3500 deg$^2$ of imaging and spectroscopy of 240,000
  stars to study the structure of our Milky Way, and {\em
    Supernovae} (Frieman \etal\ 2007), which repeatedly images a $\sim
  300$ deg$^2$ equatorial stripe in the Southern Galactic Cap to
  search for supernovae in the redshift range $0.05 < z < 0.35$ for
  measurement of the redshift-distance relation.  

  DR6 is cumulative, in the sense that it includes all data that were
  included in previous data releases.  However, as we describe in
  detail in this paper, we have incorporated into this data release a
  number of improvements and additions to the software.  These
  include:
\begin{itemize} 
\item Improved photometric calibration, using overlaps between the
  imaging scans;
\item Improved wavelength and flux calibration of the spectra;
\item Improved velocity dispersion measurements for galaxies;
\item Results of an independent determination of galaxy and quasar
  redshifts and stellar radial velocities; 
\item Effective temperatures, surface gravities and metallicities for
  many stars with spectra.
\end{itemize}
All DR6 data, including those
  included in previous releases, have been reprocessed with the new
  software.   

  In \S~\ref{sec:sky_coverage}, the sky coverage of the data included
  in DR6 is presented.  Section~\ref{sec:image} describes new features of
  the imaging data, including extensive low-latitude imaging, target
  selection of the SEGUE plates, improved
  photometric calibration, and a recently recognized systematic error
  in sky subtraction which affects the photometry of bright galaxies.
  Section~\ref{sec:spectra} describes the extensive reprocessing we
  have done of our spectra, including improved flux and wavelength
  calibration, the determination of surface temperatures,
  metallicities and gravities of stars with spectra, the
  availability of two independent determinations of object redshifts,
  and improved velocity dispersions of galaxies.  We summarize DR6 in
  \S~\ref{sec:conclusions}. 

\section{The Sky Coverage of the SDSS DR6}
\label{sec:sky_coverage}

\begin{figure}[t]\plotone{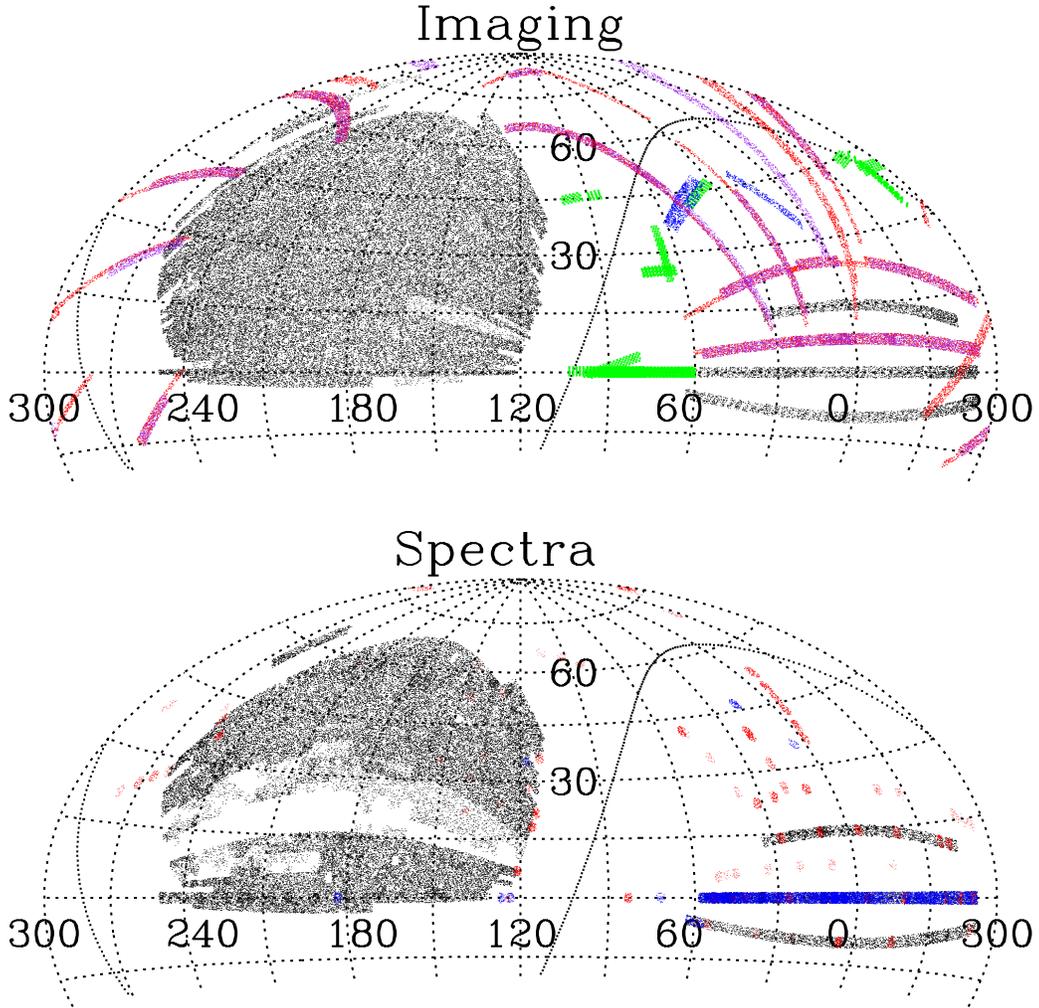}\caption{
The distribution on the sky of the data included in DR6 (upper panel:
imaging; lower panel: spectra), shown in an
Aitoff equal-area projection in J2000 Equatorial Coordinates.  The
Galactic Plane is the sinuous line that goes through each panel.  The
center of each panel is at $\alpha = 120^\circ \equiv 8^{\rm 
  h}$, and that the plots cut off at $\delta = -20^\circ$.  
The Legacy imaging survey covers the contiguous
area of the Northern Galactic Cap (centered roughly at $\alpha =
200^\circ, \delta = 30^\circ$), as well as three stripes (each of width
$2.5^\circ$) in the Southern Galactic Cap.  The regions new to DR6
are shown in lighter shading than the rest in both panels.  In addition, 
several stripes (indicated in blue in the imaging data) are auxiliary
imaging data in the vicinity of M31 and the Perseus Cluster, while the
SEGUE imaging scans are available in the DAS and CAS (red) and DAS
only (purple).  The green scans
are additional runs as described in Finkbeiner \etal\ (2004).  In the
spectroscopy panel, special plates (in
the sense of the DR4 paper) are indicated in blue, while SEGUE plates
are in red.  Note that many plates overlap; for example, there are
SEGUE plates in the contiguous area of the Northern Galactic Cap, and
the Equatorial Stripe in the Southern Galactic Cap, which appears
solid blue, is also completely covered by the Legacy survey.
\label{fig:skydist}}\end{figure}

In the Spring of 2006, the imaging for the SDSS Legacy survey was
essentially completed.  The Northern Galactic Cap in DR6 is now contiguous,
with the exception of 10 deg$^2$ spread among several holes in the
survey; these have since been imaged, and will be included in the
Seventh Data Release.  The Northern Galactic Cap imaging survey covers
7668 deg$^2$ in DR6; the additional Legacy scans in the Southern
Galactic Cap bring the total to 8417 deg$^2$.  
  The sky coverage of the
imaging data is shown in Figure~\ref{fig:skydist}, and is tabulated in
Table~\ref{table:dr6_contents}.  The images,
spectra, and resulting catalogs are all available from the DAS; with a
few exceptions noted below, all the catalogs are available from the
CAS as well.

\begin{deluxetable}{lr}
\tablecaption{Coverage and Contents of DR6
              \label{table:dr6_contents}}

\startdata

\cutinhead{\bf Imaging} 

 Imaging area in CAS &9583\ deg$^2$\\
 Imaging catalog in CAS & 287 million unique objects\\
 \quad Legacy footprint area & 8417\ deg$^2$ (5\% increment over DR5)\\
 \quad Legacy imaging catalog & 230 million unique objects \\
 \quad SEGUE footprint area, available in DAS\tablenotemark{a} & 1592\ deg$^2$\\
 \quad SEGUE footprint area, available in CAS & 1166\ deg$^2$\\
 SEGUE imaging catalog & 57 million unique objects\\
 M31, Perseus scan area & $\sim 26$ deg$^2$\\
 Southern Equatorial Stripe with $> 50$ repeat scans& $\sim 300$ deg$^2$\\
 Commissioning (``Orion'') data &832 deg$^2$\\
\cutinhead{\bf Spectroscopy}
 Spectroscopic footprint area &7425 deg$^2$ (20\% increment over DR5)\\
 \quad Legacy &6860 deg$^2$\\
 \quad SEGUE  &565 deg$^2$\\
 Total number of plate observations (640 fibers each) & 1987 \\
 \quad Legacy survey plates & 1520 \\
 \quad SEGUE plates & 162 \\
 \quad Special program plates & 226 \\
 \quad Repeat observations of plates & 79\\
 Total number of spectra& 1,271,680\\
 \quad  Galaxies\tablenotemark{b}   &  790,860 \\
 \quad  Quasars    & 103,647 \\
 \quad  Stars      & 287,071 \\
 \quad  Sky        & 68,770 \\
 \quad  Unclassifiable & 21,332\\ 
 Spectra after removing skies and duplicates & 1,115,971\\

\enddata
\tablenotetext{a}{Includes regions of high stellar density, where the
  photometry is likely to be poor.  See text for details.}
\tablenotetext{b}{Spectral classifications from the {\tt spectro1d}
  code; numbers include duplicates. The complete MAIN sample (Strauss \etal\ 2002) includes
  585,719 galaxies after duplicates are removed, while the luminous
  red galaxy sample (Eisenstein \etal\ 2001) contains 79,891 galaxies.}
\end{deluxetable}

The imaging data are the union of three data sets:
\begin{itemize} 
\item Legacy data, which includes the
large contiguous region in the Northern Galactic Cap, as well as three
$2.5^\circ$ wide stripes in the Southern Galactic Cap.  These are
shown in gray.  The lighter gray indicates those regions new to
DR6, containing 417 deg$^2$; the entire Legacy area available in DR6
is 8417 deg$^2$. 
\item Imaging stripes (also $2.5^\circ$ wide) as part of the SEGUE
  survey.  These do not aim to cover a contiguous area, but are
  separated by roughly $20^\circ$ and are designed to sparsely
  sample the large-scale distribution of stars in the Galactic halo.
  These cover just under 1600 deg$^2$, and are all available in the
  DAS.    Notice that many of these stripes go to quite low Galactic 
  latitude, and some cross the Galactic Plane.  As we describe in
  \S~\ref{sec:segue_imaging}, the SDSS photometric pipeline is not
  optimized for crowded fields, and thus the photometry of objects at
  the lowest Galactic latitudes is not reliable.  
 Of these data, 1166 deg$^2$ are
  available in the CAS in a separate database from the Legacy imaging;
  these are 
  the regions in which the outputs of the photometric pipeline are
  most reliable, and which have been used for spectroscopic targeting
  (\S~\ref{sec:segue_TS}).   The SEGUE imaging that is available in both CAS
  and DAS is indicated in red; purple indicates the area only
  available in the DAS.  
\item Additional imaging taken as part of various auxiliary programs
  as part of the SDSS, including scans of the region around M31 and
  Perseus (see the description in the DR5 paper), and adding up to
  roughly 26 deg$^2$.  These scans are
  indicated in blue.   These data are not included in the CAS,
  but are available in the DAS.  
\end{itemize}

\begin{figure}[t]\plotone{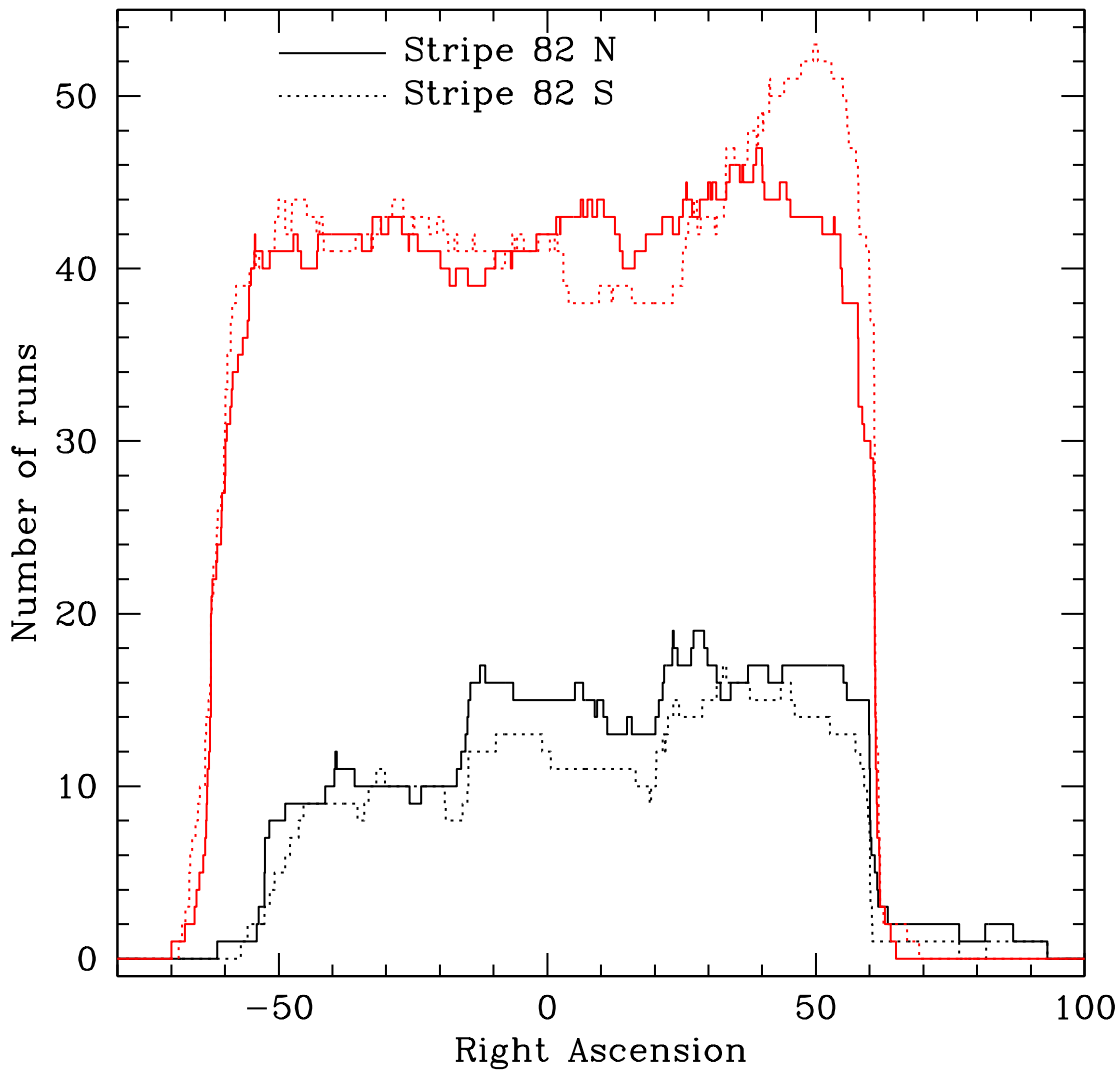}\caption{
Stripe 82, the Equatorial stripe in the South Galactic Cap, has been
imaged multiple times.  The lower pair of curves in black show the number of
scans covering a given right ascension in the North and South strip
through Fall 2004 (these data 
were also included in DR5); these data are available through the DAS.
Since that time, Stripe 82 has been covered many more times as part of
a comprehensive survey for $0.05 < z < 0.35$ supernovae, although often
in conditions of poor seeing, bright moon, and/or clouds; the number
of additional scans at each right ascension in the North and South
strip is indicated in red.  These latter data have not been flux-calibrated. 
\label{fig:stripe82}}\end{figure}

In addition, the $2.5^\circ$ wide Equatorial Stripe
(``Stripe 82'') in the Southern Galactic Cap has been imaged multiple
times through the course of the SDSS, and again as part of the
Supernova component of SDSS-II (Frieman \etal\ 2007).  Sixty-five scans of
Stripe 82 observed through Fall 2004 are of survey quality, i.e.,
they were taken under moonless and cloudless skies in good seeing.  As
in DR5, we
make the calibrated object catalogs and the images corrected for bias, flatfield,
and image defects available through the
DAS. There were an additional 171
supernova runs taken in the Fall seasons of 2005 and 2006. Much  
of these data were taken under non-photometric conditions, 
poor seeing, or during bright moon, and thus the photometry is not
reliable at face value (although Ivezi\'c \etal\ 2007 have
demonstrated that it can be calibrated quite well after the fact).
The images and the uncalibrated object catalogs for these runs
are made available through the DAS as well.  Stripe 82 is
composed of two overlapping strips (York \etal\ 2000), and
Figure~\ref{fig:stripe82} shows the number of times each right
ascension of the two strips is covered in the data through 2004 and as
part of the Supernova survey.   

Finkbeiner \etal\ (2004) made available 470 deg$^2$ of imaging
  on the Southern Equatorial Stripe taken early in the survey but not
  included in either the DAS or the CAS.  With DR6, we release an
  additional 362 deg$^2$ of imaging data; these runs are indicated in
  green in Figure~\ref{fig:skydist}.

The DR6 spectroscopy contains 1,271,680 spectra over 1987 plate observations. Of
these, 1520 plates are from the 
main Legacy survey, and there are 64 repeat observations (``extra
plates'') of 55 distinct Legacy plates.   
In addition, there are 234 observations of 226 distinct ``special"  
plates of the various programs described in the DR5 paper\footnote{An  
updated special-plate list is at \url{\tt 
{http://www.sdss.org/dr6/products/spectra/special.html}} .}, indicated
in blue in 
Figure~\ref{fig:skydist}, and 169 observations of 162 distinct  
special plates taken as part of SEGUE (see \S\ref{sec:segue_TS})
(indicated in red).  
In total, these plates cover 7425 deg$^2$ (not including overlaps).
Thirty-two
fibers (64 fibers for the SEGUE plates) are dedicated to background
sky subtraction on each plate, about 0.7\% of spectra are repeat
observations on overlapping plates for 
quality assurance (and science; see e.g., Wilhite \etal\ 2005) and
roughly 1\% of spectra are too low in signal-to-noise ratio (S/N) for
unambiguous classification, so there are roughly 1.1 million distinct
objects with useful spectra in the DR6. 
This represents a roughly 20\% increase over DR5.  The areas of sky
new to DR6 are represented in lighter gray in Figure~\ref{fig:skydist}.  We
plan to complete the spectroscopy of the contiguous area of the
Northern Galactic Cap in the Spring of 2008. 

The average seeing (see Figure 4 of the DR1 paper) and limiting magnitude of
the imaging data, as well as the typical S/N of the
main survey spectra, are essentially unchanged from previous data
releases; see the summary of survey characteristics in Table 1 of the
DR5 paper.   

\section{Characterization and use of the imaging data}
\label{sec:image}

The SDSS photometric processing pipeline has been stable since DR2,
and thus the quantities measured for all objects included in DR5 have
been copied wholesale into DR6. This version of the pipeline has
been used for the small amount of Northern Galactic Cap data new to
DR6, as well as the SEGUE imaging scans shown in
Figure~\ref{fig:skydist}.  The magnitudes quoted in the SDSS archives
are asinh magnitudes (Lupton \etal\ 1999).  

\subsection{SEGUE data at low Galactic latitudes}
\label{sec:segue_imaging}

The SEGUE imaging survey is designed to explore the structure of the
Milky Way at both high and low Galactic latitudes, and thus extends to lower
latitudes than did the Legacy survey.  This extension gives us
better leverage on the spatial 
distribution of stars in the disk components of the Milky Way, and on
the three-dimensional shape of the stellar halo.  Eighty-six of the 162
SEGUE plates were targeted off SEGUE imaging, while the remainder were 
targeted off Legacy imaging.  The SEGUE imaging scans are made
available in a separate database, termed ``SEGUEDR6'', within the CAS.

The SEGUE imaging data close to the Galactic plane have regions of
higher dust extinction and object density than does the high-latitude SDSS.
The SDSS imaging reduction pipelines used to reduce the data for DR6
were not designed for optimal performance in crowded fields, and are
known to fail for some of these data.  In particular: 
\begin{itemize} 
\item When the images are sufficiently crowded, the code has trouble
  finding suitable isolated stars from which to measure the point
  spread function (PSF).  Without a suitable determination of the PSF,
  the brightness 
  measurements by the pipeline (Stoughton \etal\ 2002) are inaccurate.
\item The pipeline attempts to deblend objects with overlapping
  images, but the deblend algorithm fails when the number of
  overlapping objects gets too large, such as 
  happens in sufficiently crowded fields.  In such fields, the number
  of detected objects reported by the pipeline can be a dramatic
  underestimate.
\item At low latitudes, the dust causing Galactic extinction (as
  measured by Schlegel, Finkbeiner \& Davis 1998, hereafter SFD) cannot be assumed to
  lie completely in front of the stars in the sample.  This has an
  effect on the interpretation of quality assurance tools based on the
  position of the stellar locus, as we describe below. 
\end{itemize}
Therefore, it is necessary to check that the
quality of the reductions in any area of the sky of interest is
adequate to address a particular science application of the data.

As Ivezi\'c \etal\ (2004) and the DR3 paper explain, we use a series
of automated quality checks on the imaging data to determine whether
the data meet our science requirements; the results of these tests are
made available in the CAS.  These checks are available for the SEGUE
imaging as well.  The best indicator of bad PSF photometry is the
difference between PSF and large aperture magnitudes for stars
brighter than 19th magnitude.  If the median difference between the
two is greater than 0.03 mag, the PSF photometry will not make the
survey requirement of 2\% calibration error in $g, r$, or $i$.  About
2.3\% of the fields of the SEGUE imaging data loaded into the CAS in
DR6 fail this criterion\footnote{Of course, a much larger fraction of
  the additional SEGUE data available in the DAS also fail this
  criterion.}.  For comparison, about 1.6\% of all fields  
in the SDSS Legacy footprint in DR6 fail this criterion.
 
The automated overall measurement of the quality in a given field also takes into
account the location of the stellar locus 
in the $ugr$ and $gri$ color-color diagrams, and how it differs in each
field from the average value over the entire survey (see the
discussion in Ivezi\'c \etal\ 2004).  These
color-color diagrams are made with SFD extinction-corrected magnitudes,
so even for very good photometry they may vary from the survey average
if that extinction correction is not valid for any reason.  The user
should apply appropriate caution in interpretation of the stellar
locus location diagnostics in the quality assurance for these data. 

Finally, the photometric pipeline performs poorly for a stellar
density greater than $\sim 5000$ objects brighter than the detection
limit per $10^\prime \times 
13^\prime$ field, or about $140,000$ objects deg$^{-2}$, a density
roughly ten times the density at high latitudes.  The outputs of the
photometric pipeline are quite incomplete (and indeed, confusingly,
can fall well below 5000 objects per field) and can be unreliable for more
crowded fields.  Almost all the SEGUE data affected by this problem
are in the DAS only; the SEGUE imaging in the CAS (which is the subset
used for SEGUE target selection; see below) largely avoids these
crowded areas of the sky.

\subsection{SEGUE target selection}
\label{sec:segue_TS}

SEGUE has as one of its goals a kinematic and stellar population study
of the high-latitude thick disk and halo of the Milky Way.  The halo
is sampled sparsely with a series of tiles each of seven deg$^2$ in both the
SEGUE imaging stripes and the main Legacy survey area, with centers
separated by roughly 10 deg.  Each such tile is sampled with two
pointings, one plate for stars brighter than $r=17.8$
(approximately the median target magnitude), and one plate,
which typically gets double the standard exposure time, for
fainter stars.  The target selection categories and criteria are
summarized in Table~\ref{table:segue_target} (listed roughly in order
from bluest to reddest targets); see the DR4 paper for a
description of an earlier version of SEGUE targeting.   Most of the
target selection categories are sparsely sampled, with a sampling rate
that depends on magnitude; see the on-line documentation for more
details. The target selection bits in the {\tt PrimTarget} flag are
indicated in the table.  Spectra with target selection bits set by
the SEGUE target selection algorithm 
have {\tt PrimTarget} bit 0x80000000 and {\tt SecTarget} bit
0x40000000 set.

Half of the science targets on each line of sight are 
selected using color-color and color-magnitude cuts designed to
sample at varying densities across the main sequence from $g-r = 0.75$
(K dwarfs at $T_{\rm
eff} < 5000$K). To this sample we
add metal-poor main sequence turnoff stars selected by their blue
$ugr$ colors, essentially an ultraviolet excess cut that is highly 
efficient at separating the halo from the thick disk near the turnoff.
At the faint end, $r=19.5$, the average star that makes this selection is at a
heliocentric distance of 10 kpc for [Fe/H]${}=-1.54$.  To reach to
greater distances, we use the strength of the Balmer jump to select
field blue horizontal branch (BHB) stars in the $ugr$ color-color
diagram (Lenz \etal\ 1998, Sirko \etal\ 2004; Clewley \etal\ 2004). 
The halo BHB sample extends to distances of 40
kpc at $g=19$ (corresponding to the S/N limit we use for detailed
spectroscopic classification; see \S~\ref{sec:stellar_parameters}). We
select all available BHB candidates in our high-latitude 
fields, and all candidates with $g-r<0$ irrespective of latitude.

We select distant halo red giant candidates by the photometric offset in
the $ugr$ color-color diagram, as quantified by the $l$ color 
(Lenz \etal\ 1998; see the notes to Table~\ref{table:segue_target}).  This offset is
caused by their ultraviolet excess and 
weak Mg I$b$ and MgH at at 5175\AA\ relative to foreground disk dwarfs (Morrison \etal\ 2001,
Helmi \etal\ 2003).  This is augmented by a 3$\sigma$ proper motion cut using a
recalibrated version of the USNO-B catalog (Munn \etal\ 2004).
Spectroscopic identification of true giants using the methodology in
Morrison \etal\ (2003) has shown that the giant selection is roughly
50\% efficient at $g<17$, the current limit to which we can reliably
distinguish giants from dwarfs in the spectra. The halo
giant sample identified in 
this way reaches distances of 40 kpc from the Sun.  We select
candidate low-metallicity stars using a more extreme $l$-color cut,
and without any proper motion cut.  

The spectroscopic selection also includes smaller categories of rare
but interesting objects.  These include cool white dwarfs selected
with the recalibrated USNO-B reduced proper motion diagram, which can
be used to date the age of the Galactic disk (Gates \etal\ 2004;
Harris \etal\ 2006), high proper motion targets from the SUPERBLINK
catalog (L\'{e}pine \& Shara 2005), which have uncovered some of the
most extreme M subdwarfs known (L\'{e}pine \etal\ 2007) and have aided
in the calibration of their metallicity scale using common proper
motion pairs, and white dwarf/main sequence binaries containing cool
white dwarfs, which are predicted to be the dominant population among
this type of binaries (Schreiber \& G\"ansicke 2003).  These rare
object categories also include color-only selections for cool
subdwarfs, brown dwarfs (using cuts similar to those employed by Chiu
\etal\ 2006), and the SEGUE ``AGB" category that selects 
metal-rich, cool giants that separate readily from the $ugr$ stellar
locus.

Table~\ref{table:segue_target} describes Version 4\_2 of the SEGUE
target selection algorithms. The algorithms have evolved throughout the 
survey, and users wishing to understand the detailed selection
associated with each target category should examine the SEGUE
documentation off the DR6 survey page.  The user should also know that
SEGUE target selection has been run only on those chunks used to
design SEGUE plates, and has not yet been run on the bulk of the
Legacy survey imaging.

\begin{deluxetable}{p{3cm}p{2cm}p{9cm}r}
\tablecaption{SEGUE targeting algorithms\label{table:segue_target}}
\tablehead{\colhead{Category}&\colhead{Bit (Hex)}&\colhead{Color
    cuts}&\colhead{\#/tile}}
\startdata
White dwarf & 0x80080000 &     
$g < 20.3, -1 < g-r < -0.2, -1 < u-g < 0.7, u-g + 2(g-r) < -0.1$ &25\\
A, BHB stars &0x80020000&
    $g < 20.5$, $0.8 < u-g < 1.5$, 
    $-0.5 < g-r < 0.2$&$\le$155\\
Metal-poor MS turnoff& 0x80100000&
    $g < 20.3$,
    $-0.7 < P_1 < -0.25$,
    $0.4 < u-g < 1.4$,
    $0.2 < g-r < 3.0$
 & 200 \\
F/G stars &0x80000200&
$14.0 < g < 20.2$,    $0.2 < g-r < 0.48$&50\\
G stars& 0x80040000&
    $14.0 < r < 20.2$, 
    $0.48 < g-r < 0.55$&375\\
Cool$\,$white$\,$dwarf & 0x80020000&
    $14.5 < r < 20.5$, $-2 < g-i$, 
$ H_g > \max[17.5, 16.05+2.9(g-i)]$,
   $$g-i < \left\{ \begin{array}{ll}
0.12 & \mbox{neighbor with $g
    < 22$ within $7''$}\\
    1.7 &\mbox{otherwise}
		   \end{array}
\right. $$
    &10\\
Low metallicity& 0x80010000&
    $r < 19.5$, 
    $-0.5 < g-r < 0.75$,
    $0.6 < u-g < 3.0$, 
    $l > 0.135$&150\\
K giant& 0x80040000&
    $r < 20.2$,
    $0.7 < u-g < 4.0$, 
    $0.5 < g-r < 0.9$, 
    $0.15 < r-i < 0.6$, 
    $l > 0.07$,
    $\mu < 0.011^{\prime\prime}$/yr&95\\
K dwarf& 0x80008000&
    $14.5 < r < 19.0$, 
    $0.55 < g-r < 0.75$&95\\ 
MS/WD pairs &0x80001000&
    $15 < g < 20$,
    $u-g < 2.25$,
    $-0.2 < g-r < 1.2 $,
    $0.5 < r-i < 2.0$,
    $ -19.78(r-i) + 11.13 < g-r < 0.95(r-i) + 0.5 $,
    $$i-z > \left\{
     \begin{array}{rl}
       0.5 &\mbox{if $r-i > 1.0$} \\
       0.68(r-i)-0.18 &\mbox{otherwise}
       \end{array}
       \right.$$
&5-10\\
M subdwarf&0x80400000&
    $14.5 < r < 19.0$,
    $g-r > 1.6$, 
    $0.95 < r-i < 1.3$
&5\\
High $\mu$ M subdwarf& 0x80400000&
$\mu > 0.04^{\prime\prime}$/yr,  $r-z > 1.0$, $15 + 3.5 (g-i) > H_r > 12 + 3.5
(r-z)$&60\\
Brown dwarf&0x80200000&
    $z < 19.5 $,
    $u > 21 $,
    $g > 22 $,
    $r > 21 $,
    $i-z > 1.7$
&$<$5\\
AGB &0x80800000&
    $14.0 < r < 19.0$, 
    $2.5 < u-g < 3.5$,
    $0.9 < g-r < 1.3$, 
    $s < -0.06$&10\\ 
\enddata
\tablecomments{
This table describes Version 4\_2 of the SEGUE target selection
algorithm.  The hex bit in the second column is set in the {\tt PrimTarget} flag.  All
magnitudes above are PSF magnitudes which have been corrected for 
Galactic extinction following SFD. The one exception is the MS/WD pair
algorithm, which uses PSF magnitudes without extinction correction.
The quantity $l \equiv -0.436u + 1.129g-0.119r -0.574i+0.1984$ is a metallicity indicator following 
Lenz \etal\ (1998).  The quantities $s \equiv -0.249u+0.794g-0.555r+0.234$
and $P_1 \equiv 0.91(u-g) + 0.415(g-r) -1.280$
are defined by Helmi \etal\ (2003).  The proper motion $\mu$ is in
units of arcsec/yr, and the reduced proper motion is defined as $H_g
\equiv g + 5\log \mu +5$ and similarly for $H_r$.  The fourth column
lists the typical number of targets selected in each category per
spectroscopic tile.}
\end{deluxetable}

\subsection{A caveat on high proper motion stars}
\label{sec:highpm}
As described in the DR2 paper, the proper motions of stars in the SDSS
are taken from the measurements of the 
USNO-B1.0 (Monet \etal\ 2003), based primarily on the POSS-I and
POSS-II.  However, this catalog is incomplete at the highest proper
motions, greater than 100 milli-arcsec per year.  Confusingly, objects
with no proper motion measurement in the USNO-B1.0 catalog have their
proper motion listed as 0.0 in the CAS {\tt ProperMotions} table, meaning
that a query for 
{\em low} proper motion stars will be contaminated by a small number
of the highest proper motion stars.  The best available catalog of
high proper motion stars can be found in the SUPERBLINK catalog of
L\'epine \& Shara (2005) and references therein; we hope to
incorporate this catalog into the proper motion data in the SDSS in
future data releases.   

\subsection{Low Galactic latitude SDSS commissioning data}
\label{sec:orion}

During commissioning and subsequent tests of the SDSS observing system,
additional data were obtained outside of the nominal survey region.
These data consist of 28 runs (see Finkbeiner \etal\ 2004, Table 1)
at low Galactic latitude, mostly in the star-forming regions of Orion,
Cygnus, and Taurus.  There are 832 deg$^2$ of data, 470 deg$^2$ of which
have been previously released\footnote{At
\url{{\tt http://photo.astro.princeton.edu}} .} as flat files.
There are three types of files: 
calibrated images (one \texttt{calibImage} per field), calibrated
object files (one \texttt{calibObj} per field), and condensed ``sweep"
files (one star or galaxy file per run/camcol).

The remaining 362 deg$^2$ are hereby released in
the same format, but they are not available in the DAS or CAS.  These
data have been photometrically calibrated using the \"ubercalibration
algorithm (\S~\ref{sec:ubercal})\footnote{The current
  \"ubercalibration has yielded calibrations typically 0.02
mag different from those previously released, but some runs/camera
columns show differences as large as 0.05 mag. The variance within
each field is also somewhat reduced by correcting flatfield errors at
the 0.01 or 0.02 mag level.}.  \"Ubercalibration takes
advantage of the Apache Wheel calibration scans (not shown in
Figure~\ref{fig:skydist}) to tie the photometry of disjoint regions of
the sky together; nevertheless, because the overlap with other runs is
less than in the main survey area, their calibration may not be quite
as good.  

\subsection{Improved photometric calibration}
\label{sec:ubercal}
Photometric calibration in SDSS has been carried out in two parallel
approaches.  The first uses an auxiliary $20''$ photometric telescope
(PT) at the site, which 
continuously surveys a series of US Naval Observatory standard 
stars which are used to define the SDSS $u'g'r'i'z'$ photometric system 
(Smith \etal\ 2002). Transformations between the $u'g'r'i'z'$ and native 
SDSS 2.5-meter $ugriz$ photometric systems and zeropoints for stars in 
patches surveyed by the 2.5-meter telescope are determined with these 
data (Tucker \etal\ 2006, Davenport \etal\ 2007). These
secondary patches are spaced roughly 
every $15^\circ$ along the imaging stripes.  This approach has allowed
the SDSS photometry to reach its goals of calibration errors with an
rms of 2\% in $g$, $r$, and $i$, and 3\% in $u$ and $z$ (Ivezi\'c
\etal\ 2004), as measured from repeat scans (see the discussion in
Ivezi\'c \etal\ 2007).   This is the calibration process that has been
used in all data releases to date.  However, it is not ideal for
several reasons: 
\begin{itemize} 
\item The $u'g'r'i'z'$ filter system of the PT camera is subtly
  different from the $ugriz$ system on the 2.5-meter;
\item There are persistent problems with the flat-fielding of both the PT
 and 2.5-meter cameras, especially in $u'$;
\item No use is made of overlap data in the 2.5-meter scans to tie
  the zeropoints together.  
\end{itemize}

A second approach, termed ``\"ubercalibration'' 
(Padmanabhan \etal\ 2007) does not use information from the PT to
calibrate individual runs, but rather uses the overlaps between the
2.5-meter imaging runs to tie the photometric zeropoints of individual
runs together and measure the 2.5m flatfields, and to determine the
extinction coefficients on each night.  Unlike the standard PT
calibrations, \"ubercalibration explicitly 
assumes that the photometric calibration parameters -- a zeropoint
for each CCD, and atmospheric extinction linear with airmass -- are  
constant through a photometric night.This assumption appears justified, as
the resulting calibration has errors of $\sim 1\%$ in $g, r, i$ and
$z$, and $2\%$ in $u$, roughly a factor of two below
those of the standard processing, as determined from the overlaps
themselves, and from the measurement of the ``principal colors'' of the stellar
locus (see the discussion in Ivezi\'c \etal\ 2004 and the DR3 paper).  This
scatter is dominated by unmodelled variations in the atmospheric
conditions in the site, including changes in the atmospheric
extinction through a night.  

The
relative calibration of the photometric scans via overlaps does not
determine the photometric zeropoints in the five filters.  The
zeropoints are constrained in practice by forcing the \"ubercalibrated photometry
of bright stars to agree in the mean with that calibrated in the
standard way (Tucker \etal\ 2006).  Thus this 
work does not represent an 
improvement in the calibration of the SDSS photometry to a true AB
system (in which magnitudes can be translated directly into physical
flux units); see the discussion in the DR2 paper, Eisenstein
\etal\ (2006), and Holberg \& Bergeron (2006).  Moreover, there are
subtle differences between the response of the six filters in each row of
the SDSS camera, especially in $z$ (see the discussion in Ivezi\'c
\etal\ 2007); these differences have not been corrected. 

  Both versions of SDSS photometry are now made available through the
  CAS in DR6.  The PT-calibrated photometry for each detected object
  is stored in the database in the same tables and columns as in DR5,
  and both the offset between PT and \"ubercalibration, as well as the
  \"ubercalibrated magnitudes, are stored in the {\tt UberCal} table
  of the CAS. Database functions are available to apply these offsets
  and output \"ubercalibrated photometry.  The distribution of these
  offsets is shown in Figures 15 and 16 of Padmanabhan \etal\ (2007);
  the improvements are subtle, changing magnitudes of most individual
  objects by 0.02 mag or less.

\subsection{The photometry of bright galaxies}
\begin{figure}[t]\plotone{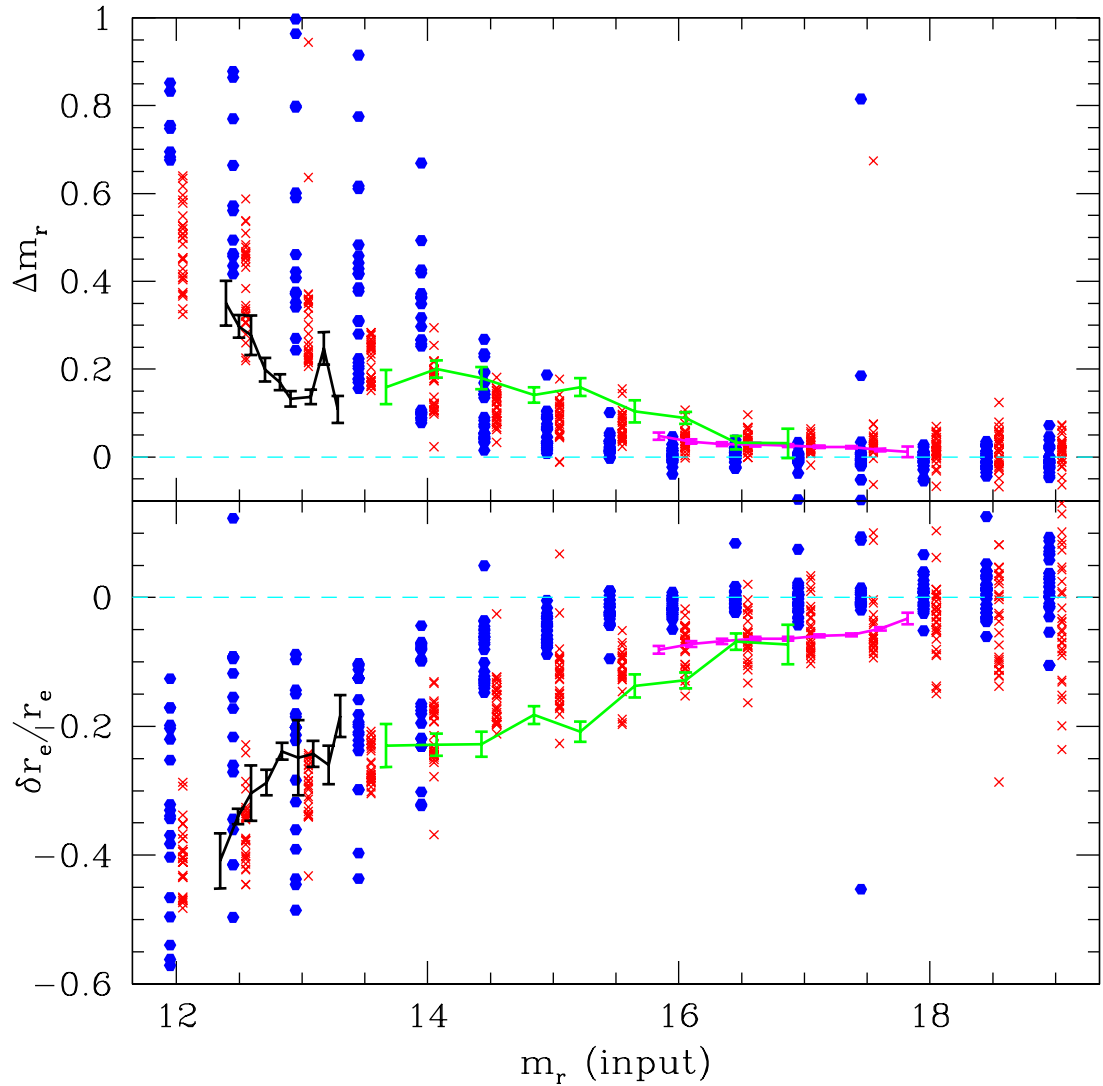}\caption{
The effects of sky subtraction errors on the photometry of bright
galaxies. 
{\it Upper panel:} The error in the $r$ band model magnitude of simulated galaxies with
an $n=1$ (exponential) profile (blue hexagons) and an $n=4$ (de
Vaucouleurs) profile (red crosses) as determined by the photometric
pipeline, as a function of magnitude.  Fifteen galaxies are simulated
at each magnitude for each profile.  Also shown are the analogous
results from Hyde \& Bernardi (unpublished) for three early-type
galaxy samples: 54
nearby ($z<0.03$) early-type galaxies from the ENEAR catalog (da Costa
\etal\ 2000) in black; 280 
brightest cluster galaxies from the C4 catalog (Miller \etal\ 2005) in
green; and 9000 early-type galaxies from the Bernardi \etal\ (2003a)
analysis in magenta.   
{\it Lower panel:} The fractional error in the scale size $r_e$ as a
function of magnitude from the simulations and the Hyde \& Bernardi
analysis.   
\label{fig:mandelbaum}}\end{figure}

Because of scattered light (see the EDR paper), the background sky in
the SDSS images is non-uniform on arc-minute scales.  The photometric
pipeline determines the median sky value within each $101.4''$ (256
pixel) square on a grid with $50.7''$ spacing, and bilinearly interpolates this sky value to each
pixel.  This procedure overestimates the sky near large extended
galaxies and bright stars, and as 
was already reported in the DR4 paper and Mandelbaum \etal\ (2005),
causes a systematic decrease in the number density of faint objects
near bright galaxies.  In addition, it also strongly affects the
photometry of bright galaxies themselves, as has been reported by
Lauer \etal\ (2007), Bernardi \etal\ (2007), and 
Lisker \etal\ (2007). 
We have quantified this
effect by adding simulated galaxies with exponential and 
de Vaucouleurs (1948) profiles to SDSS images, following Blanton \etal\ (2005a). The
simulated galaxies ranged from apparent magnitude $m_r=12$ to 
$m_r=19$ in half-magnitude steps, with a one-to-one mapping from $m_r$
to S\'ersic half-light radius determined using the mean observed relation
between these quantities for MAIN sample 
galaxies (Strauss \etal\ 2002) with exponential and de Vaucouleurs
profiles.  Axis ratios of $0.5$ 
and $1$ were used, with random position angles for the
non-circular simulated galaxies.  The results in
the $r$ band are shown in Figure~\ref{fig:mandelbaum}, plotting the
difference between the input magnitude and the model magnitude
returned by the SDSS photometric pipeline as a function of magnitude.
Also shown is the fractional 
error in the scale size $r_e$.  The biases are significant to $r=16$
for late-type galaxies, and to $r=17.5$ for early-type galaxies.  Hyde \&  
Bernardi (unpublished) fit de Vaucouleurs models to SDSS images of
extended elliptical galaxies, using their own sky subtraction
algorithm, which is less likely to overestimate the sky level near
extended sources.  Their results, also shown in the figure, are quite
consistent with the simulations.

The scatter in the offset from one realization to another is large
enough that we cannot recommend a deterministic correction for this
problem.  This scatter depends in part on the position of the
simulated galaxy relative to the grid on which the sky interpolation
occurs.  We are working on an improved algorithm which will fit
the extended profiles of galaxies explicitly as part of the sky
determination, and hope to include the results in a future data
release.

\section{Spectroscopy}
\label{sec:spectra}

The Sixth Data Release contains a number of improvements and additions
to the SDSS spectroscopy. These include an
improved pipeline to extract and calibrate the one-dimensional spectra
(\S~\ref{sec:idlspec2d}), the results of an independent pipeline to
classify objects and measure redshifts (\S~\ref{sec:specBS}), the
results of a pipeline to determine the effective temperatures, 
gravities and metallicities of stars
(\S~\ref{sec:stellar_parameters}), and improvements to the existing
code to measure velocity dispersions (\S~\ref{sec:veldisp}). 

\subsection{The extraction and calibration of one-dimensional spectra}
\label{sec:idlspec2d}
The pipeline that extracts, combines, and calibrates the SDSS spectra
of individual objects from the two-dimensional spectrograms
(``{\tt idlspec2d}'') was
originally designed to obtain meaningful redshifts for galaxies and
quasars.  However, there were several ways in which the code was
inadequate, especially in light of the stellar focus of the SEGUE
project, and the recognition of the rich stellar data available among
the spectra of the main SDSS survey.  The spectrophotometry was tied
to the fiber magnitudes of stars, whose relation to the true, PSF
magnitudes of stars is seeing-dependent.  In addition, the SEGUE
spectroscopy includes ``bright plates'' which contain substantial
numbers of stars as bright as $i_{fiber} = 14.2$, and scattered light
from these stars caused systematic errors in the sky subtraction on
these plates.  Finally, there were errors in the wavelength
calibration as large as 15 km s$^{-1}$ on some plates, acceptable for
most extragalactic science, but a real limitation for SEGUE's science
goals. These 
concerns and others have caused us to substantially revise and improve
the {\tt idlspec2d} pipeline; the results of this improvement are included
in DR6.  

\subsubsection{Spectrophotometry: Flux Scale}
  The new code has a different spectrophotometric calibration flux
  scale.  The fiber magnitude 
  reported by the photometric pipeline is the brightness of each
  object, as measured through a $3''$ diameter aperture corrected to
  $2''$ seeing to match the entrance aperture of the fibers (see the
  discussion in the EDR paper).  However, the relationship between the
  fiber magnitudes of stars and the PSF magnitudes (which, for
  unresolved objects, is our best determination of a true, total
  magnitude) is dependent on seeing; this is made worse because the
  {\em colors} of stars measured via fiber magnitudes will be
  sensitive to the different seeing in the different filters (although
  cases in which the seeing is dramatically different in the different
  bands are fairly rare).  With
  this in mind, the pipeline used in DR6 determines the spectrophotometric
  calibration on each plate such that the flux of the spectrum of
  standard stars integrated over the filter curve matches the PSF
  magnitude of the stars as measured from their imaging.  This
  calibration is determined for each of the four cameras (two in each 
  spectrograph) from observations of standard stars.  Additional
  corrections to handle large-scale astrometric and chromatic terms
  are measured from isolated stars and galaxies of high S/N, and are
  then applied to all the objects on the plate.  

\begin{figure}[t]\plotone{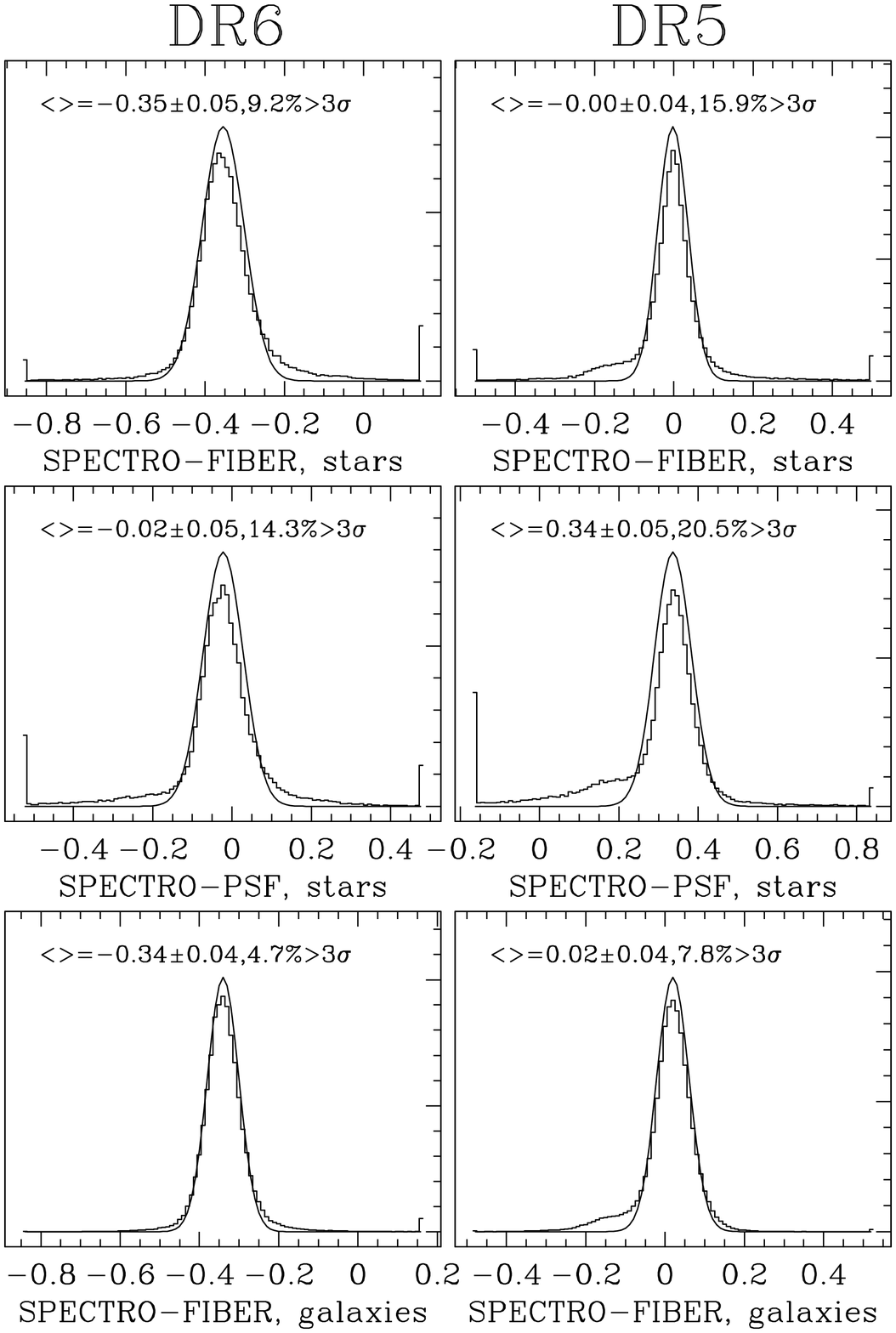}\caption
{The distribution of differences between $r$-band photometry synthesized from
SDSS spectra (labelled ``SPECTRO''), and PSF and fiber magnitudes, for
stars and galaxies; 
results are shown for DR6 (left-hand panel) and the previous version
of the calibration available in DR5 (right-hand panel).  Only objects
with PSF magnitude brighter 
than 19 are shown.  The
most important difference is the offset of 0.35 magnitudes between the
two, due to the change in calibration from fiber to PSF photometry.
Each panel includes the mean and standard deviation of the best-fit
Gaussian, as well as the number of objects lying beyond $3\sigma$ (as
a measure of the non-Gaussianity of the tails).  Results are shown for
$r$ band, but $g$ and $i$ band results are very similar. 
\label{fig:spectrophotometry}}\end{figure}

  The results of this calibration may be seen in
  Figure~\ref{fig:spectrophotometry}, which compares synthesized
  magnitudes from the SDSS spectra with the PSF and fiber magnitudes
  in the imaging data, showing results both from the old (``DR5'') and
  new (``DR6'') codes.  We emphasize that
  the calibration is {\em not} tied to the PSF photometry of each
  object individually (otherwise the comparison in
  Figure~\ref{fig:spectrophotometry} would be a tautology); there is a
  single calibration determined for each camera in a given plate.
  This means, for example, that it is meaningful to compare photometry
  and spectrophotometry of individual objects to look for variability
  (e.g., Vanden Berk \etal\ 2004).

  The PSF includes light that extends beyond the $3''$ diameter of the
  filters, and thus the PSF-calibrated spectrophotometry is
  systematically {\em brighter} than the old fiber-calibrated
  photometry by the difference between PSF and fiber magnitudes, which
  is roughly 0.35 magnitudes (albeit dependent on seeing).  Again,
  because the PSF photometry represents an accurate measure of the
  brightness of stars, this calibration means that the
  spectrophotometry matches the PSF photometry for stars to an rms of 
  4\%. 
  This distribution does show an extended tail presumably
  caused by blended and variable objects\footnote{Indeed, the fiber
    magnitudes include light from overlapping blended objects, thus
    the tails are 
    less extensive in the fiber magnitude comparison.}, but the distribution is substantially
  more symmetric than for the previous version of the pipeline.  
  Interestingly, for galaxies, the rms difference between spectroscopic
  photometry and the fiber magnitudes is also 4\%.  The
  previous code shows a similarly narrow distribution, albeit with
  larger tails.  The distribution of the difference of the $g-r$ and
  $r-i$ colors between PSF photometry and as 
  synthesized from the spectrophotometry again shows a 
  narrow core in both DR5 and DR6, but again with less extensive
  non-Gaussian outliers with the new code. 

  Due to errors in the processing step, there are 28 plates, listed in
  Table~\ref{table:fiber_mag_calibrated_plates}, that were calibrated
  using fiber magnitudes rather than PSF magnitudes.  Therefore,
  objects on these plates have a spectroscopic flux scale
  systematically lower by 0.35 mag than the rest of the survey. These
  will be processed correctly in a subsequent data release.

\begin{deluxetable}{ll ll ll ll}
\tablecaption{Spectroscopic plates calibrated with fiber
  magnitudes\label{table:fiber_mag_calibrated_plates}}
\tablehead{\colhead{plate}&\colhead{MJD}&\colhead{plate}&\colhead{MJD}&\colhead{plate}&\colhead{MJD}&\colhead{plate}&\colhead{MJD}}
\startdata
269&51910&345&51690&460&51924&683&52524\\ 
270&51909&349&51699&492&51955&730&52466\\ 
277&51908&353&51703&543&52017&830&52293\\ 
284&51943&367&51997&554&52000&872&52339\\ 
309&51666&394&51913&556&51991&1394&53108\\
324&51666&403&51871&616&52374&1414&53135\\
336&51999&446&51899&616&52442&1453&53084\\
\enddata
\tablecomments{The second column lists the Modified Julian Date (MJD)
  on which each plate was observed.}
\end{deluxetable}

\subsubsection{Spectrophotometry: Wavelength Dependence}
\begin{figure}[t]\plotone{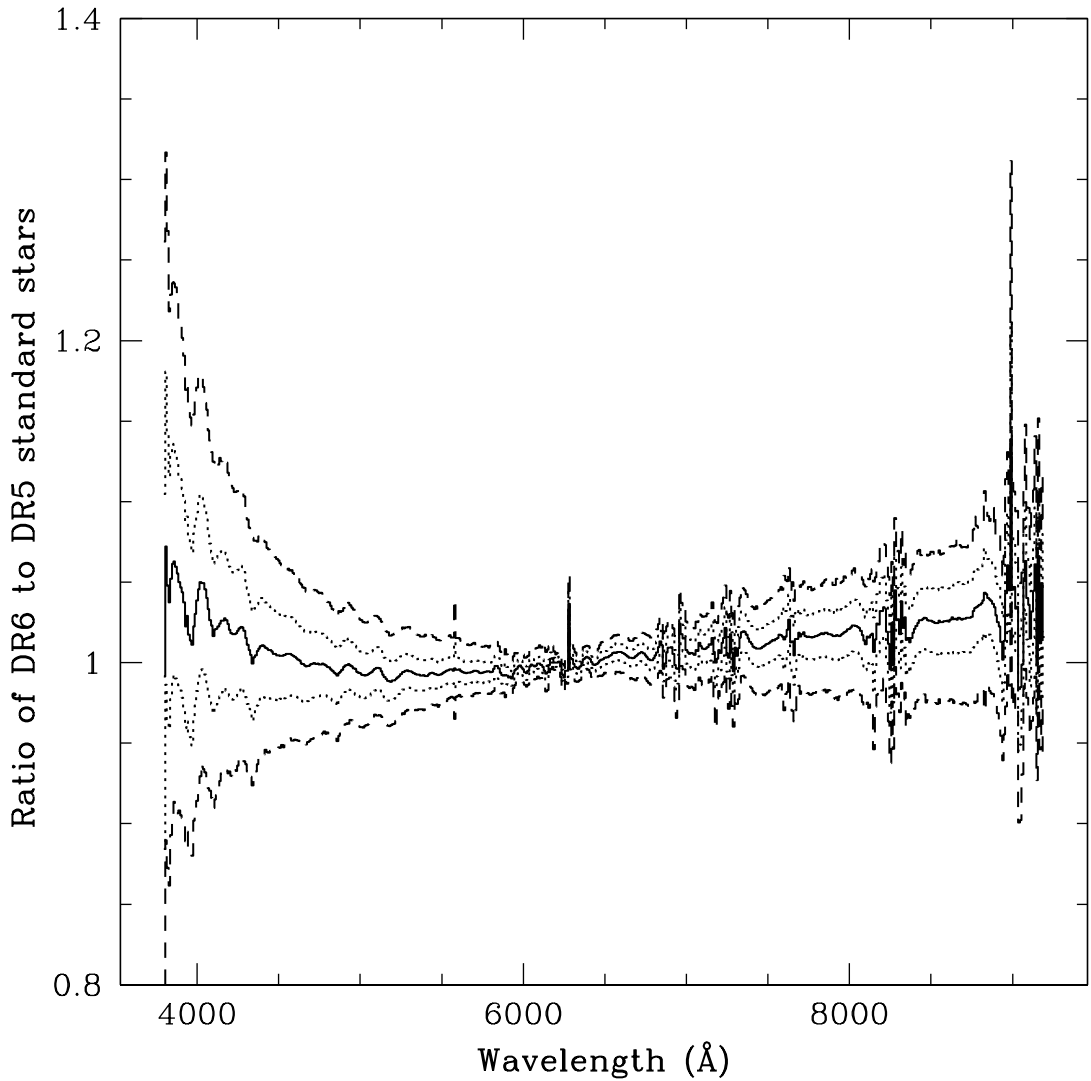}\caption{
The ratio of the summed spectra of standard stars on
  each plate as determined by the DR6 and DR5 versions of
  spectrophotometry, rescaled to unity at 6200\AA.  The solid line is
  the median ratio spectrum over 1278 plates, the dotted lines
  enclose 68.3\% of the plates (corresponding to 1$\sigma$ for a
  Gaussian distribution), and the dashed lines enclose 95.4\% of the
  plates (corresponding to 2$\sigma$).  The distribution at each
  wavelength is in fact close to Gaussian.  
\label{fig:compare_tremonti}}\end{figure}

As discussed in the DR2 paper, each plate includes observations of a
number of spectrophotometric standards, typically F subdwarfs.  Their
observed spectra are fit to and calibrated against the models of Gray
\& Corbally (1994), as updated by Gray \etal\ (2001).  We can compare
the spectrophotometric calibration between DR5 and DR6 by plotting the
ratio of the summed spectra of these standard stars on each plate as
determined by the two versions of the pipeline.  The 0.35 mag overall flux
scale between the two calibrations has been taken out by forcing all
the curves through unity at 6200\AA.  The median ratio (as determined
from 1278 plates), and the 68.3\% and 95.4\% outliers, are shown in
Figure~\ref{fig:compare_tremonti}. The median ratio differs from unity
by less than 5\% at all wavelengths, but a small fraction of the
plates have differences as large as 30\% at the far blue end.

\begin{figure}[t]\plotone{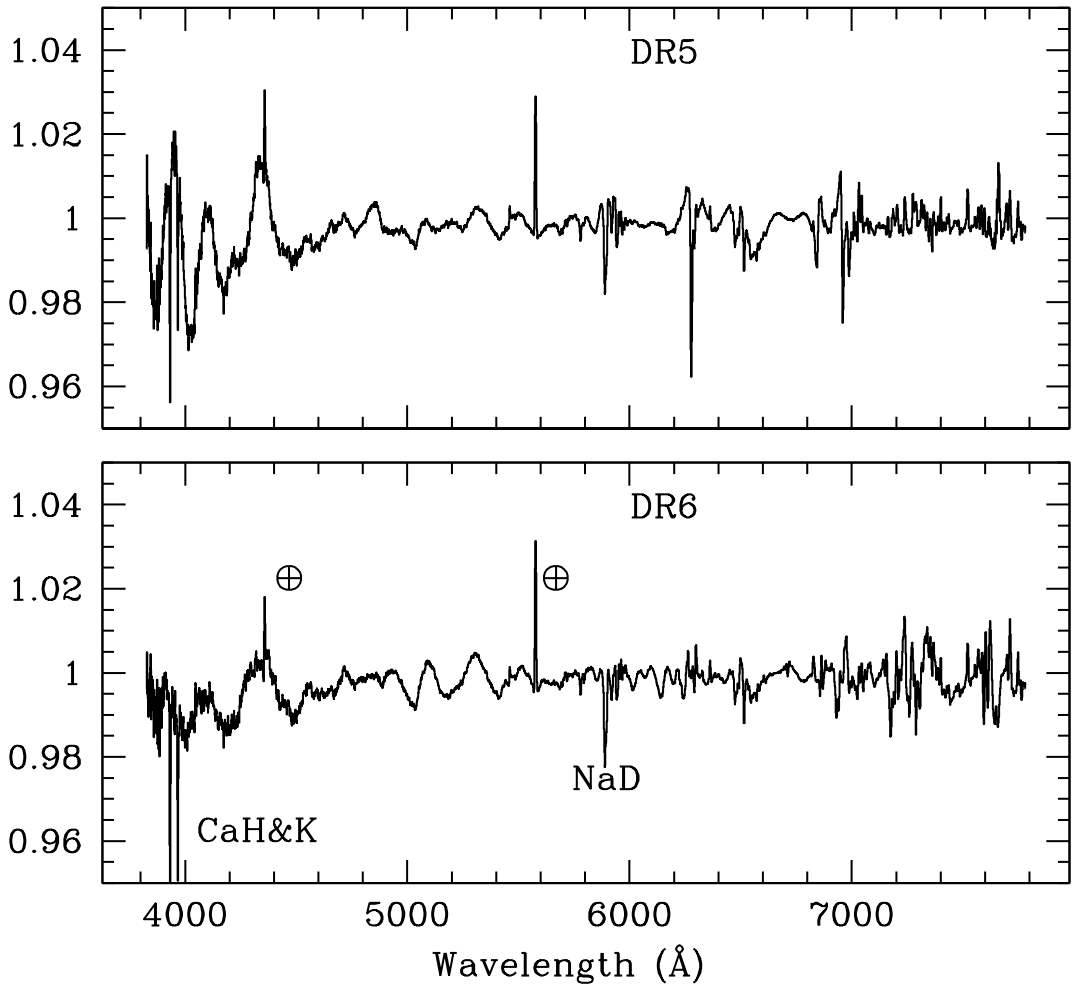}\caption{The
    median ratio of observed flux-calibrated spectra of luminous red
    elliptical galaxies to their averaged spectra (after taking
    evolution into account), for the previous (DR5) and current (DR6)
    spectroscopic reductions.  This quantifies the
    wavelength dependence of systematic errors in the
    spectrophotometric calibration; the amplitude of these features,
    already small in the previous reductions, have been reduced
    further in DR6, especially in the blue.  The features at Ca H and
    K and at Na D are probably due to absorption from the interstellar
    medium.  The strong features at the sky lines at 5577\AA\ and
    4358\AA\ marked with the $\oplus$ symbol are related to the S/N of
    the spectra; a similar analysis 
    with quasar spectra shows these features to have substantially
    lower amplitude.
\label{fig:spectro_residuals}}\end{figure}

  Do these changes represent an improvement scientifically?  Figure 4
  of the DR2 paper quantified the uncertainties in the 
  spectrophotometric calibration used at that time by looking at the
  mean fractional offset between observed spectra of white dwarfs and
  best-fit models for them.  Figure~\ref{fig:spectro_residuals} shows
  a similar analysis with the old and new reductions.  The
  curves show the median fractional difference between a sample of
  128,000 calibrated
  luminous red galaxy (LRG, Eisenstein \etal\ 2001) spectra, 
and a
  model based on averaged observed LRG spectra that is 
  allowed to evolve smoothly with redshift (see the discussion in \S~3
  of the DR5 paper).
  Because the LRGs have a broad range of redshifts, one expects no
  feature specific to the LRGs to appear in this plot as a function of
  observed wavelength, and deviations from unity are a measure of the
  small-scale errors in the spectrophotometry.  There are systematic
  oscillations at the 2\% level in the DR5 reductions. These wiggles
  correspond to positions of strong absorption lines in
  the standard stars, especially in the vicinity of the 4000\AA\ break
  in the blue.  This is now handled by not fitting the instrumental
  response to any residual non-telluric features finer than 25-50\AA,
  as the response is not expected to vary on those scales.  This
  reduces the amplitude of the wiggles by a factor of two in the DR6
  reductions, especially at the blue end.  Redward of 4500\AA, 50\% of
  the spectra fall within 3\% of the median value; this increases to
  7\% at 3800\AA.   
The
  features at Ca K and H (3534 and 3560\AA) and Na D (5890 and 5896\AA) are
  probably due to absorption from the interstellar medium (although
  the latter probably has a contribution from sky line residuals).  The sky
  line residuals (marked with the $\oplus$ symbol) are a function of
  S/N; a similar analysis with higher S/N quasars shows substantially
  smaller residuals at the strong sky lines.

\begin{figure}[t]\plotone{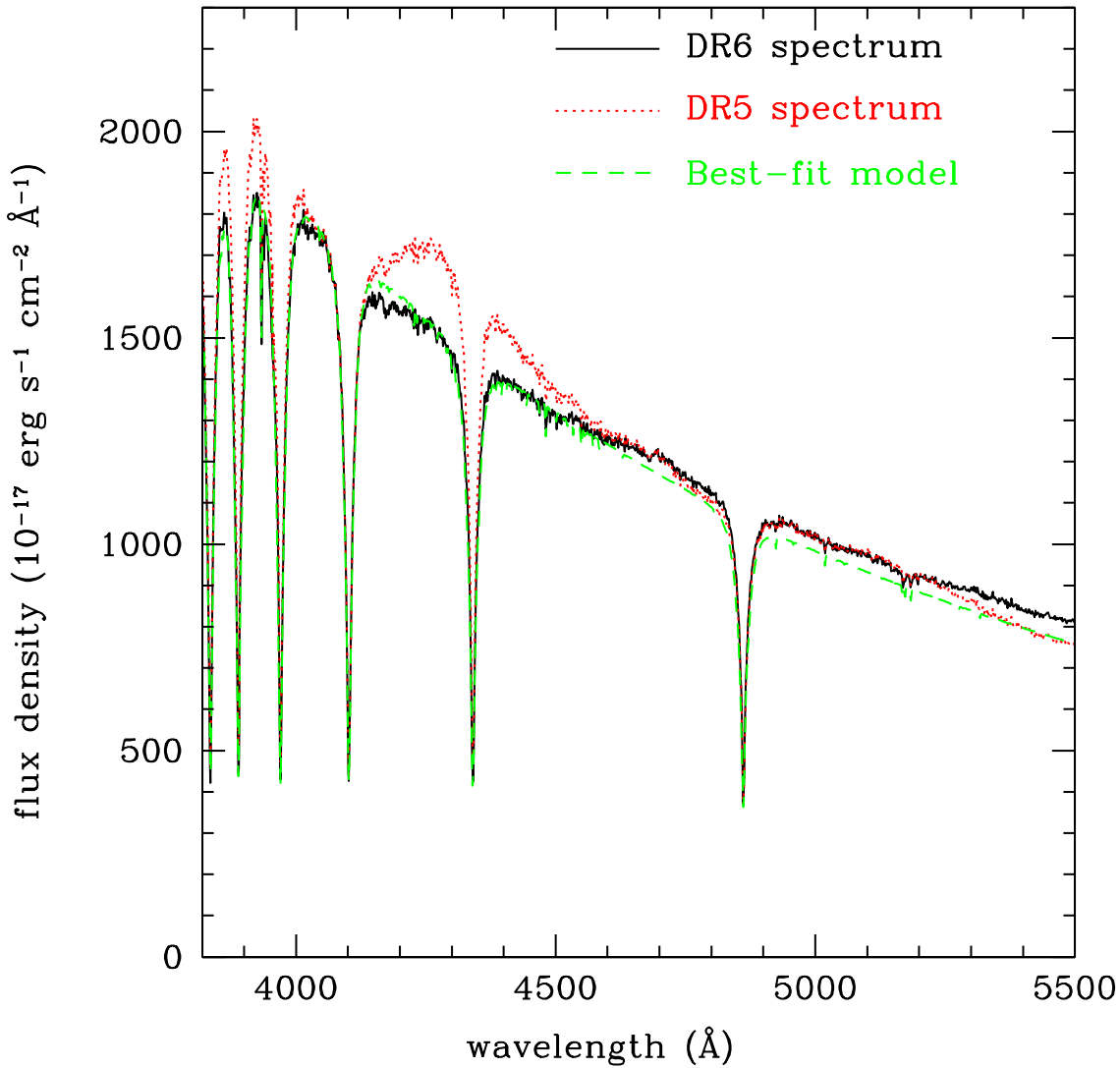}
\caption{The blue part of the spectrum of an A0 blue horizontal branch
  star, SDSS J004037.41+240906.5, as given by the old (red dotted
  curve) and new (black solid curve) versions of {\tt idlspec2d}.  The
  old curve has been scaled up 
  to reflect the difference in the calibration of the two reductions.
  The synthetic spectrum, shown in green, is generated from a model
  with parameters matching those derived from the {\tt SSPP} ($T_{\rm eff} =
  8500~K$, $\log g = 3.25$, [Fe/H] = $-2.00$).  The continuum between
  the absorption lines is much smoother, and matches the synthetic
  spectrum much better for the new reductions than for the old.  The
  synthetic spectrum has been normalized to match the observed spectrum
  at 4500\,{\AA}.  Neither the model nor the spectra have been
  corrected for Galactic reddening (which is $E(B-V)=0.036$ in this
  line of sight).  
\label{fig:harding}}\end{figure}

The effect of this improvement in the spectrophotometric calibration
becomes clear if we examine the spectra of individual stars.
Figure~\ref{fig:harding} shows the blue part of the spectrum of an A0
blue horizontal branch star as calibrated with the old code (dotted)
and the new (solid), together with a synthetic spectrum based on the
atmospheric parameters estimated by the SEGUE Stellar Parameter
Pipeline (\S~\ref{sec:stellar_parameters}; $T_{\rm eff} = 8446~K$,
$\log g = 3.15$, [Fe/H] = $-1.96$).  The new reductions are clearly
smoother between the absorption lines; the match between the DR6
calibrated spectrum and the synthetic spectrum is also superior.

\subsubsection{Radial velocities}
In order to measure the dynamics of the halo of the Milky Way, SEGUE
requires stellar radial velocities accurate to 10 km s$^{-1}$, 
significantly more demanding than the original SDSS requirements of 30
km s$^{-1}$.  The previous version of {\tt idlspec2d} had systematic
errors of 10--15 km s$^{-1}$ in the wavelength calibration because of
a dearth of strong lines at the blue end of the spectrum in the
calibration lamps and in the nighttime sky.  The sky-line
fits for the blue side wavelength corrections now use a more robust
algorithm allowing less freedom in the fits, and these problems are
largely under control.   

We monitor the systematic and random errors in the radial velocities
in the SEGUE data by comparing repeat observations on the bright and
faint plates of each SEGUE pointing.  The duplicate observations
consist of roughly 20 ``quality assurance" objects selected at the
median magnitude of the SEGUE data, as well as a similar number of
spectroscopic calibration objects that are observed on both plates.
The mean difference in the measured radial velocities between the two
observations of the quality assurance objects depends on stellar type,
with a standard deviation of 9 km s$^{-1}$ for A and F stars and 5 km
s$^{-1}$ for K stars\footnote{Thus the error on a single star is
  $\sqrt{2}$ less than these values.}. The mean radial velocity offset
between the two plates in each pointing, as measured using all the
duplicate observations, suggests systematic velocity errors from plate to plate of
only 2 km s$^{-1}$ rms. 

We have checked the zeropoint of the overall radial velocity scale (as
measured using the
ELODIE templates in the {\tt specBS} code; see the discussion below in
\S~\ref{sec:specBS}) by carrying
out high-resolution observations of 150 SEGUE stars.  This
has revealed a systematic error of 7.3 km s$^{-1}$ (in the sense that
the SpecBS velocities are too low) due to subtly different
algorithms in the line fits to arc and sky lines. This has been fixed
in the output files of the {\tt SSPP} (\S~\ref{sec:stellar_parameters}
below), but has {\em not} yet been fixed elsewhere in the CAS.

The improved wavelength calibration leads to smaller sky subtraction
residuals for many objects, especially noticeable in the far red of
the spectrum.

\subsubsection{Additional outputs}
Under good conditions, a typical spectroscopic plate is observed three
times in exposures of 15 minutes each; more exposures are added in
poor conditions to reach a target S/N in the spectra.  The {\tt idlspec2d} pipeline stitches
together the resulting individual spectra to determine the final spectrum of
a given object.  However, for the most accurate determination of the
noise characteristics of the spectra (for example, in detailed
analyses of the Lyman $\alpha$ forest of quasars; see the discussion
in McDonald \etal\ 2006), or to determine whether a specific unusual
feature in a spectrum is real, it is desirable to go back to the
uncombined spectra.  These uncombined spectra are now made available
for every plate in the so-called {\tt spPlate} files through the DAS. 

The published spectra have had a determination of the spectrum of the
foreground sky subtracted from them.  The sky is measured in 32 fibers
(64 fibers for the faint SEGUE plates) 
placed in regions where no object has been detected to 5$\sigma$ in
the imaging 
data, interpolated (both in amplitude and in wavelength, allowing for
some undersampling) to each
object exposure, and subtracted.  However, it is often useful to see
the sky spectrum that has been subtracted from each object, for
example to study the nature of extended foreground emission-line
objects in the data (for example, see Hewett \etal\ 2003 for the discovery of a
$2^\circ$ diameter planetary nebula in the SDSS data).  The sky
spectrum subtracted from each object spectrum is now available 
through both the DAS and the CAS.  

\subsubsection{The treatment of objects with very strong emission
  lines}

There is a known problem, which is not fixed with the current version
of {\tt idlspec2d}, whereby the code that combines the individual
15-minute exposures will occasionally mis-interpret the peaks of
particularly strong and narrow emission lines as cosmic rays and
remove them.  All pixels affected by this have the inverse variance
(i.e., the inverse square of the estimated error at this pixel) set to
zero, indicating that the code recognizes that the pixel in question
is not valid.  A diagnostic of this problem is unphysical line ratios
in the spectra of dwarf starburst galaxies, as the tops of the
strongest lines are artificially clipped.  This is a rare problem,
affecting less than 1\% of galaxies with rest equivalent width in
the H$\beta$ line greater than 25\AA, but users investigating the
properties of galaxies with strong emission lines should be aware of
it.  We hope to fix this problem in the next data release.  

\subsection{An independent determination of spectral classifications
  and redshifts}
\label{sec:specBS}

As described in the EDR paper and Subbarao \etal\ (2002), the spectral
classifications and radial velocities available in the data releases
have been based on a code ({\tt spectro1d}), that cross-correlates the observed spectra
with a variety of templates in Fourier space to determine
absorption-line redshifts and fits Gaussians to emission lines to
determine emission-line redshifts.  A completely
separate code, termed {\tt specBS} and written by D. Schlegel (in
preparation) instead carries out $\chi^2$ fits of the spectra to
templates in wavelength space (in the spirit of Glazebrook \etal\ 1998),
allowing galaxy and quasar spectra to be fit with linear combinations
of eigenspectra and low-order polynomials.  Stellar radial
velocities are fit both to SDSS-derived stellar templates, and to
templates drawn from the high-resolution ELODIE 
(Prugniel \& Soubiran 2001) library.  
The {\tt spectro1d} outputs give the default spectroscopic information
available through the CAS, but the {\tt specBS} outputs are made
available through the CAS for the first time with DR6\footnote{The
  outputs of {\tt specBS} have also been made 
  publically available through the NYU Value-Added Galaxy Catalogue;
  see Blanton \etal\ (2005b).}.  While {\tt spectro1d} uses manual
inspection to correct the redshifts and classifications of a small
fraction of its redshifts, {\tt specBS} is completely automated.  

Tests show that the two pipelines give impressively consistent
results.  At high S/N, the rms difference between the redshifts of the
two pipelines is of order 7 km s$^{-1}$ for stars and galaxies,
although the {\tt spectro1d} redshifts are systematically higher by 12
km s$^{-1}$ due to differences in the templates.  The difference
distribution has non-Gaussian tails, but as a test of catastrophic
errors, we find that 98\% of {\em all} objects with spectra (after
excluding the blank sky fibers) have consistent classification (star,
quasar, galaxy) and redshifts agreeing within 300 km s$^{-1}$ for
galaxies and stars, and 3000 km s$^{-1}$ for quasars.

Half of the remaining 2\% are objects of very low S/N, and the other
half are a mixture of a variety of unusual 
objects, including BL Lacertae objects (Collinge \etal\ 2005; their
lack of spectral features makes it unsurprising that the two pipelines
come to different conclusions), unusual white dwarfs, including strong
magnetic objects and metal-rich systems (Schmidt \etal\ 2003; Eisenstein \etal\ 2006;
Dufour \etal\ 2007), unusual broad absorption
line quasars (Hall \etal\ 2002), superposed objects, including at
least one gravitational lens (Johnston \etal\ 2003), and so on.
Both pipelines set flags when the classifications or
redshifts are uncertain (see Table~\ref{table:specBS-warning}); the
majority of these discrepant cases are flagged as uncertain by both
pipelines.

Table~\ref{table:specBS} lists the outputs from the {\tt specBS} pipeline
included in the CAS for each object.  In addition, the DAS includes
the results of the cross-correlation of each of the templates with
each spectrum, as well as Gaussian fits to the emission lines.  These
quantities are included in the {\tt SSPP} table
(\S~\ref{sec:stellar_parameters}) in the CAS, and as flat files in the
DAS.

\begin{deluxetable}{llp{10cm}}
\tabletypesize{\small}
\tablecaption{Redshift warning flags from {\tt specBS}}
\tablehead{\colhead{Bit}&\colhead{Name}&\colhead{Comments}}
\startdata
0 & SKY\_FIBER &Fiber is used to determine sky; there should be no
object here.\\
1 & SMALL\_LAMBDA\_COVERAGE &Because of masked pixels, less than half
of the full wavelength range is reliable in this spectrum.\\ 
2 & CHI2\_CLOSE & The second best-fitting template had a reduced
$\chi^2$ within 0.01 of the best fit (common in low S/N spectra).\\
3 & NEGATIVE\_TEMPLATE & Synthetic spectrum is negative (only set for stars and QSOs).\\
4 & MANY\_5SIGMA & More than 5\% of pixels lie more than $5\,\sigma$ from
the best-fit template.\\
5 & CHI2\_AT\_EDGE & $\chi^2$ is minimized at the edge of the
redshift-fitting region (in this circumstance, Z\_ERR is set to
$-1$).\\
6 & NEGATIVE\_EMLINE & A quasar emission line (C IV, C III], Mg II,
  H$\beta$, or H$\alpha$) appears in absorption with more than
  $3\,\sigma$ significance due to negative eigenspectra.\\
\enddata
\label{table:specBS-warning}
\end{deluxetable}

\begin{deluxetable}{lr}
\tabletypesize{\small}
\tablecaption{Outputs of the {\tt specBS} pipeline made available in the DR6
  CAS.}
\tablehead{\colhead{Parameter}&\colhead{Comments}}
\startdata
CLASS   &STAR, GALAXY, or QSO\\
SUBCLASS &Stellar subtype, galaxy type (starforming, etc)\\
Z       &Heliocentric redshift\\
  Z\_ERR &Error in redshift\\
  RCHI2  &Value of reduced $\chi^2$ for template fit to spectrum\\
  DOF    &Degrees of freedom in $\chi^2$ fit\\
  VDISP  &Velocity Dispersion for galaxies (km s$^{-1}$) \\
  VDISP\_ERR &Error in Velocity Dispersion (km s$^{-1}$) \\
  ZWARNING &Set if the classification or redshift are uncertain; see
  Table~\ref{table:specBS-warning}\\
  ELODIE\_SPTYPE &Spectral type of best-fit ELODIE template\\
  ELODIE\_Z &Redshift determined from best-fit ELODIE template\\
  ELODIE\_Z\_ERR &Error in redshift determined from best-fit ELODIE template\\
\enddata
\label{table:specBS}
\end{deluxetable}

 \subsection{The measurement of stellar atmospheric parameters from the spectra}
\label{sec:stellar_parameters}

The SEGUE science goals require accurate determinations of effective
temperature, $T_{\rm eff}$, surface gravity ($\log g$, where $g$ is in
cgs units, cm s$^{-2}$), and metallicity
[Fe/H]), for the stars with spectra (and $ugriz$ photometry) obtained
by SDSS. We have developed the SEGUE Stellar
Parameter Pipeline ({\tt SSPP}), to determine these quantities and measure
77 atomic and molecular line indices for each 
object. 
The code and its performance is described in detail by Lee
\etal\ (2007a).  Validation of the sets of parameters based on
Galactic open and globular clusters and with high-resolution
spectroscopy obtained for over 150 SDSS/SEGUE stars is discussed by
Lee \etal\ (2007b) and Allende Prieto \etal\ (2007).  Due to the wide
range of parameter space covered by the stars that are observed, a
variety of techniques are used to estimate the
atmospheric parameters; a decision tree is implemented to decide which
methods or combination of methods provide optimal measures, based on
the colors of the stars and S/N of the spectra. 

These methods include:

\begin{itemize}

\item Fits of the spectra to synthetic photometry and
  continuum-corrected spectra based on Kurucz (1993) model atmospheres
  (Allende Prieto \etal\ 2006), or to synthetic spectra computed with
  the more recent Castelli \& Kurucz (2003) models (Lee \etal\ 2007a);

\item Measurements of the equivalent widths of various metal-sensitive
  lines, including the Ca~II K line (Beers \etal\ 1999) and the Ca~II
  infrared triplet (Cenarro \etal\ 2001);

\item Measurements of the equivalent widths of various
  gravity-sensitive lines such as Ca~I $\lambda$4227\AA\ and the
  5175\AA\ Mg I$b$/MgH complex (e.g.,
  Morrison \etal\ 2003);

\item Measurements of the autocorrelation function of the spectrum,
  which is useful for high-metallicity stars (Beers \etal\ 1999);

\item A neural network technique which takes the observed spectrum as
  input, trained on previously available parameters from the {\tt SSPP} (Re Fiorentin
\etal\ 2007).

\end{itemize}

For stars with temperatures between 4500~K and 7500~K and with average S/N per
spectral pixel greater than 15, the typical formal errors returned by
the code are 
$\sigma(T_{\rm eff})$ = 150~K, $\sigma(\log g)$ = 0.25 dex, and $\sigma([\rm
Fe/\rm H])$ = 0.20 dex. Comparison with 150 stars with high S/N high resolution
spectra (and therefore reliable stellar parameters) validates these error
estimates, at least for those stars with the highest quality SDSS
spectra.

The {\tt SSPP} assumes solar abundance ratios when quoting
metallicities, with the caveat that several of the individual
techniques (those that involve the Ca and Mg line strengths) adopt a
smoothly increasing [$\alpha$/Fe] ratio, from 0.0 to +0.4, as inferred
metallicity decreases from solar to [Fe/H]${}= -1.5$.  Other
techniques, which are based on regions of the spectra dominated by
lines from unresolved Fe-peak elements, do not assume such
relationships.

The S/N limit for acceptable estimated stellar parameters varies with
each individual method employed by the {\tt SSPP}.  As a general rule, the
{\tt SSPP} sets a conservative criterion that the average S/N per pixel over
the wavelength range 3800-6000\AA\ must be greater than 15 for stars
with $g-r < 0.3$, and greater than 10 for stars with $g-r \ge 0.3$.
Stars of low S/N do not have their parameters reported by {\tt SSPP}.  
Table 5 of Lee \etal\ (2007a) describes the valid ranges of effective
temperature, $g-r$ color, and S/N for each method used in the {\tt SSPP}.

The {\tt SSPP} values are combined with the outputs of {\tt specBS}
(\S~\ref{sec:specBS}) and are loaded as a single table into the CAS,
with entries for every object with a spectrum.  

For the coolest stars, measuring precise values of $T_{\rm eff}$,
$\log g$, and [Fe/H] from spectra dominated by broad molecular
features becomes extremely difficult (e.g., 
Woolf \& Wallerstein 2006).  As a result, the SEGUE {\tt SSPP} does not estimate atmospheric
parameters for stars with $T_{\rm eff} < 4500$ K, but instead
estimates the MK spectral type of each star using the {\tt Hammer}
spectral typing software developed and described by 
Covey \etal\ (2007)\footnote {The {\tt Hammer} code has been made available for
  community use: the IDL code can be downloaded from \url{\tt
    http://www.cfa.harvard.edu/$\sim$kcovey/thehammer} .}.  The {\tt
  Hammer} code measures 28 spectral indices, including atomic lines
(H, Ca I, Ca II, Na I, Mg I, Fe I, Rb, Cs) and molecular bandheads (G
band, CaH, TiO, VO, CrH) as well as two broad-band color ratios.  The
best-fit spectral type of each target is assigned by comparison to the
grid of indices measured from more than 1000 spectral type standards
derived from spectral libraries of comparable resolution and coverage
(Allen \& Strom 1995; Prugniel \& Soubiran 2001; Hawley \etal\ 2002;
Bagnulo \etal\ 2003; Le Borgne \etal\ 2003; Valdes \etal\ 2004;
S\'anchez-Bl\'azquez \etal\ 2006).  These indices, and the best-fit type
from the {\tt Hammer} code, are available for stars of type F0 and
cooler in DR6.

Tests of the accuracy of the {\tt Hammer} code with degraded (S/N $\sim$ 5)  
STELIB (Le Borgne \etal\ 2003), MILES (S\'anchez-Bl\'azquez \etal\ 2006)  
and SDSS (Hawley \etal\ 2002) dwarf template spectra reveal that the  
{\tt Hammer} code assigns spectral types accurate to within $\pm 2$ subtypes for  
K and M stars.  The {\tt Hammer} code can return results for warmer stars, but  
as the index set is optimized for cool stars, typical uncertainties  
are $\pm 4$ subtypes for A--G stars at S/N $\sim$ 5; in this temperature  
regime, {\tt SSPP} atmospheric parameters are a more reliable indicator of
$T_{\rm eff}$. 

Given SEGUE's science goals, we emphasize two limitations to the  
accuracy of spectral types derived by the {\tt Hammer} code:

\begin{itemize} 
\item The {\tt Hammer} code uses spectral indices derived from dwarf standards;  
spectral types assigned to giant stars will likely have larger, and  
systematic, uncertainties.
\item The {\tt Hammer} code was developed in the context of SDSS'
  high latitude   
spectroscopic program; the use of broad-band color ratios in the  
index set will likely make the spectral types estimated by the  
{\tt Hammer} code particularly sensitive to reddening.  Spectral types derived  
in areas of high extinction (i.e., low-latitude SEGUE plates) should be  
considered highly uncertain until verified with reddening-insensitive  
spectral indices.
\end{itemize}

\subsection{Correction of biases in the velocity dispersions}
\label{sec:veldisp}

\begin{figure}[t]\plotone{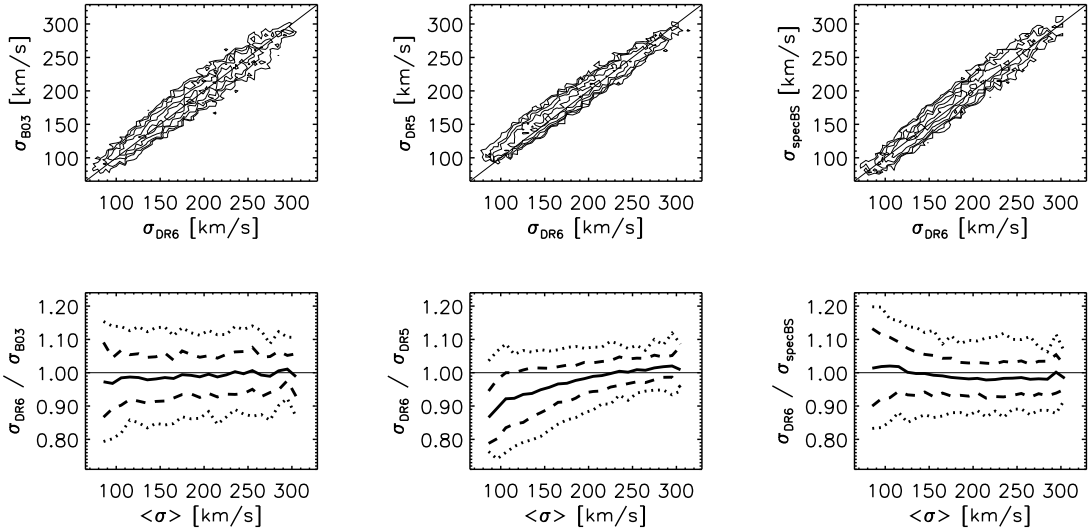}\caption{Top
    panels: velocity dispersion 
    measurements from B03 (left), 
DR5 (middle) and {\tt specBS} (right) versus the {\tt spectro1d} DR6
values for the sample of elliptical galaxies used in 
Bernardi \etal\ (2003a).  
Bottom panels: The ratio of DR6 values to the other three samples (i.e.
B03, DR5, and {\tt specBS}) versus the mean value (e.g. left panel
$\langle\sigma\rangle =(\sigma_{\rm DR6} + \sigma_{\rm
  B03})/2$). The median value at at each value of $\langle
\sigma\rangle$ is shown as a solid line; the values including 68\% and
95\% of the points are given as the dashed and dotted lines,
respectively. \label{fig:vdisp}}\end{figure}

Both {\tt specBS} and {\tt spectro1d} measure velocity dispersions
($\sigma$) for galaxies.  {\tt specBS} does so, as described above, by
including it as a term in the direct $\chi^2$ fit of templates to
galaxies.  The velocity dispersion in {\tt spectro1d} was computed as
the average of the {\it Fourier-} and {\it direct-fitting} methods
(Appendix B of Bernardi \etal\ 2003b; hereafter B03). However, due to
changes in the spectroscopic reductions from the EDR to later
releases, a bias appeared in the recent values available in the
CAS. As shown in Appendix A of Bernardi (2007), $\sigma$ values in the
DR5 do not match the values used by B03.  The difference is small but
systematic, with {\tt spectro1d} DR5 larger than B03 at $\sigma\le
150$~km~s$^{-1}$.  A similar bias is seen when comparing {\tt
  spectro1d} DR5 with measurements from the literature (using the
HyperLeda database; Paturel \etal\ 2003).  Simulations similar to
those in B03 show that the discrepancy results from the fact that the
Fourier-fitting method is biased by $\sim 15\%$ at low dispersions
($\sim 100 $~km~s$^{-1}$), whereas the direct-fitting method is not.
We therefore use only the direct-fitting method in DR6.
Figure~\ref{fig:vdisp} shows comparisons of the {\tt spectro1d} DR6
velocity dispersions with those from B03, DR5 and the {\tt specBS}
measurements. There is good agreement between {\tt spectro1d} DR6 and
B03 (rms scatter $\sim 7.5\%$), and between {\tt spectro1d} DR6 and
{\tt specBS} (rms scatter $\sim 6.5\%$), whereas {\tt spectro1d} DR5
is clearly biased high at $\sigma \le 150$~km~s$^{-1}$.  The agreement
between {\tt spectro1d} DR6 and {\tt specBS} is not surprising, since
both are now based only on the direct-fitting method.  The {\tt
  specBS} measurements tend to be slightly smaller than DR6 at $\sigma
\le 100$~km~s$^{-1}$; {\tt specBS} is similarly smaller than
HyperLeda, whereas DR6 agrees with HyperLeda at these low dispersions.

\begin{figure}[t]\plotone{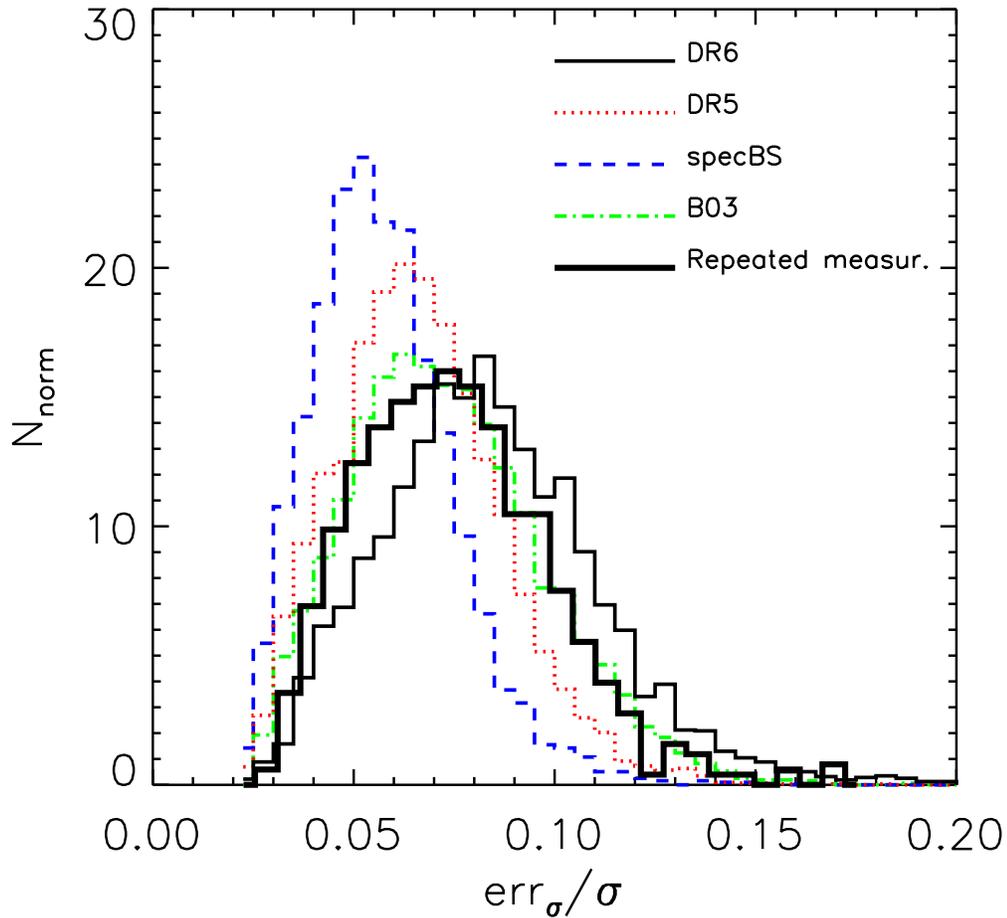}\caption{Error
    distribution of the velocity dispersion measurements from {\tt
      spectro1d} DR6 (thin black solid line), {\tt spectro1d} DR5
    (dotted red line), {\tt specBS} (dashed blue line), and B03
    (dotted-dashed green line). The thick solid line was obtained by
    comparing the velocity dispersions of $\sim 300$ galaxies which
    had been observed two or more times; it is thus an empirical
    estimate of the true error.  
    \label{fig:vdisp_err}}\end{figure}

Figure~\ref{fig:vdisp_err} shows the distribution of the error on the
measured velocity dispersions.  The direct-fitting method used by {\tt
  spectro1d} gives slightly larger errors than does the
Fourier-fitting method, peaking at $\sim 10\%$.  The figure shows that
this error distribution is consistent with that found by comparing the
velocity dispersions of $\sim 300$ objects from DR2 which had been
observed more than once.

Finally, HyperLeda reports substantially larger velocity dispersions
than SDSS at $\sigma\ge 250$~km~s$^{-1}$.  The excellent agreement
between three methods (direct fitting, cross-correlation, and
Fourier-fitting) applied to the SDSS spectra at the high velocity dispersion
end gives us confidence in our velocity dispersions (Bernardi 2007),
although it is unclear why the literature values are higher.

\subsection{Linking SEGUE imaging and spectroscopy}
For the Legacy imaging, there exist simple links between the
spectroscopic and imaging data, but these links are not yet in place
between all the SEGUE spectroscopy and imaging.  In particular, to
obtain BEST or ubercalibrated stellar photometry of SEGUE
spectroscopic objects within CAS, one must perform an ``SQL join''
command between the spectroscopic {\tt specobjall} or {\tt sppParams}
tables in the CAS with the corresponding photometric tables ({\tt
  photoobjall}, {\tt seguedr6.photoobjall}, or {\tt ubercal}). 
Sample queries on how to do this
 	are provided on the SDSS web site. 
 	We plan to simplify this procedure in future data releases.

\section{Conclusions and the Future}
\label{sec:conclusions}

We have presented the Sixth Data Release of the Sloan Digital Sky
Survey.  It includes 9583 deg$^2$ of imaging data, including a
contiguous area of 7668 deg$^2$ of the Northern Galactic Cap. The data
release includes almost 1.3 million spectra selected over 7425 deg$^2$
of sky, representing a 20\% increment over the previous data release.
This data release includes the first year of data from the SDSS-II, and
thus includes extensive low-latitude imaging data, and a great deal of
stellar spectroscopy.  New to this data release are:
\begin{itemize} 
\item 1592 deg$^2$ of imaging data at lower Galactic latitudes, as
  part of the SEGUE survey, of which 1166 deg$^2$ are in searchable
  catalogs in the CAS;
\item Revised photometric calibration for the imaging data, with
  uncertainties of 1\% in $g, r, i$ and $z$, and 2\% in $u$;
\item Improved wavelength and flux calibration of spectra;
\item Detailed surface temperatures, metallicities, and gravities for
  stars.
\end{itemize}

The SDSS-II will end operations in Summer 2008, at which point the
Legacy project will have completed spectroscopy for the entire
contiguous area of the Northern Cap region now covered by imaging,
and SEGUE will have obtained spectra for 240,000 stars.  The supernova survey
(Frieman \etal\ 2007) has discovered 327 spectroscopically confirmed SNIa to date in its
first two seasons, and has one more season to go.  

\acknowledgements

Funding for the SDSS and SDSS-II has been provided by the Alfred P. Sloan 
Foundation, the Participating Institutions, the National Science Foundation, 
the U.S. Department of Energy, the National Aeronautics and Space 
Administration, the Japanese Monbukagakusho, the Max Planck Society, 
and the Higher Education Funding Council for England. The SDSS Web Site 
is \url{{\tt http://www.sdss.org/}}.

The SDSS is managed by the Astrophysical Research Consortium for the 
Participating Institutions. The Participating Institutions are the 
American Museum of Natural History, Astrophysical Institute Potsdam, 
University of Basel, University of Cambridge, Case Western Reserve University, 
University of Chicago, Drexel University, Fermilab, the Institute for 
Advanced Study, the Japan Participation Group, Johns Hopkins University, 
the Joint Institute for Nuclear Astrophysics, the Kavli Institute for 
Particle Astrophysics and Cosmology, the Korean Scientist Group, 
the Chinese Academy of Sciences (LAMOST), Los Alamos National Laboratory, 
the Max-Planck-Institute for Astronomy (MPIA), the Max-Planck-Institute 
for Astrophysics (MPA), New Mexico State University, Ohio State University, 
University of Pittsburgh, University of Portsmouth, Princeton University, 
the United States Naval Observatory, and the University of Washington.

\end{document}